\documentclass[aip,amsmath,amssymb,reprint]{revtex4-1}

\usepackage{graphicx}
\usepackage{dcolumn}
\usepackage{bm}
\usepackage[caption=false]{subfig}

\DeclareMathOperator{\erf}{erf}
\DeclareMathOperator{\erfc}{erfc}

\begin{document}

\preprint{AIP/123-QED}

\title[Hydrodynamic Interactions in Ion Transport]{Hydrodynamic Interactions in Ion Transport -- Theory and Simulation}

\author{Diddo Diddens}
\email{d.diddens@fz-juelich.de}
\affiliation{Helmholtz-Institute M\"unster: Ionics in Energy Storage (IEK-12), Forschungszentrum J\"ulich GmbH, Corrensstra\ss{}e 46, 48149 M\"unster}
\author{Andreas Heuer}
\email{andheuer@uni-muenster.de}
\affiliation{Helmholtz-Institute M\"unster: Ionics in Energy Storage (IEK-12), Forschungszentrum J\"ulich GmbH, Corrensstra\ss{}e 46, 48149 M\"unster}
\affiliation{Institut f\"ur physikalische Chemie, Westf\"alische Wilhelms-Universit\"at M\"unster, Corrensstra\ss{}e 28/30, 48149 M\"unster}

\date{\today}

\begin{abstract}
We present a hydrodynamic theory describing pair diffusion in systems with periodic boundary conditions, thereby generalizing earlier work on self-diffusion [D\"unweg and Kremer, \textit{J. Chem. Phys.} \textbf{1993}, 99, 6983-6997; Yeh and Hummer, \textit{J. Phys. Chem. B} \textbf{2004}, 108, 15873-15879]. 
Its predictions are compared to Molecular Dynamics simulations for a liquid carbonate electrolyte and two ionic liquids, for which we characterize the correlated motion between distinct ions. 
Overall, we observe good agreement between theory and simulation data, highlighting that hydrodynamic interactions universally dictate ion correlations. 
However, when summing over all ion pairs in the system to obtain the cross-contributions to the total cationic or anionic conductivity, the hydrodynamic interactions between ions with like and unlike charges largely cancel. 
Consequently, significant conductivity contributions only arise from deviations from a hydrodynamic flow field of an ideal fluid, that is, from the local electrolyte structure as well as from relaxation processes in the subdiffusive regime. 
In case of ionic liquids, the momentum-conservation constraint additionally is vital, which we study by employing different ionic masses in the simulations. 
Our formalism will likely also be helpful to estimate finite-size effects of the conductivity or of Maxwell-Stefan diffusivities in simulations. 
\end{abstract}


\maketitle

\section{Introduction}
\label{sec:intro}

Due to the increasing demand for renewable energies, substantial efforts are currently made to develop novel electrolytes for energy storage devices \cite{xu2004nonaqueous,xu2014electrolytes}. 
For contemporary lithium ion batteries, liquid carbonate-based electrolytes remain important as well-established materials which can deliberately be fine-tuned via additives \cite{xu2004nonaqueous,xu2014electrolytes}. 
On the other hand, ionic liquids (ILs) are a more novel class of materials that are promising for e.g. supercapacitors, as they are solely composed of cations and anions and hence have high charge densities \cite{gebresilassie2014energy,watanabe2017application}. 

With respect to the application of these materials as electrolytes, the ionic conductivity 
\begin{equation}
 \label{eq:sigtot}
 \sigma = \sigma_+ + \sigma_-\text{,}
\end{equation}
containing contributions from both cations ($\sigma_+$) and anions ($\sigma_-$), as well as the transference numbers
\begin{equation}
 t_\pm = \frac{\sigma_\pm}{\sigma_+ + \sigma_-}
\end{equation}
of cations and anions, respectively, are key design parameters. 
Most generally, experimental measurements of $\sigma_\pm$ and $t_\pm$ can achieved by electrophoretic NMR (eNMR) \cite{gouverneur2015direct,brinkkotter2018relevance,hosseinioun2019improved,rosenwinkel2020coordination,pfeifer2021quantification,nurnberg2022superionicity} or combined techniques with additional assumptions \cite{zugmann2011measurement,wohde2016li+,vargas2020dynamic}. 
For polymer electrolytes, the well-established Bruce-Vincent method is another viable option \cite{bruce1988conductivity,zugmann2011measurement}. 

\begin{figure}
\includegraphics[width=0.4\textwidth]{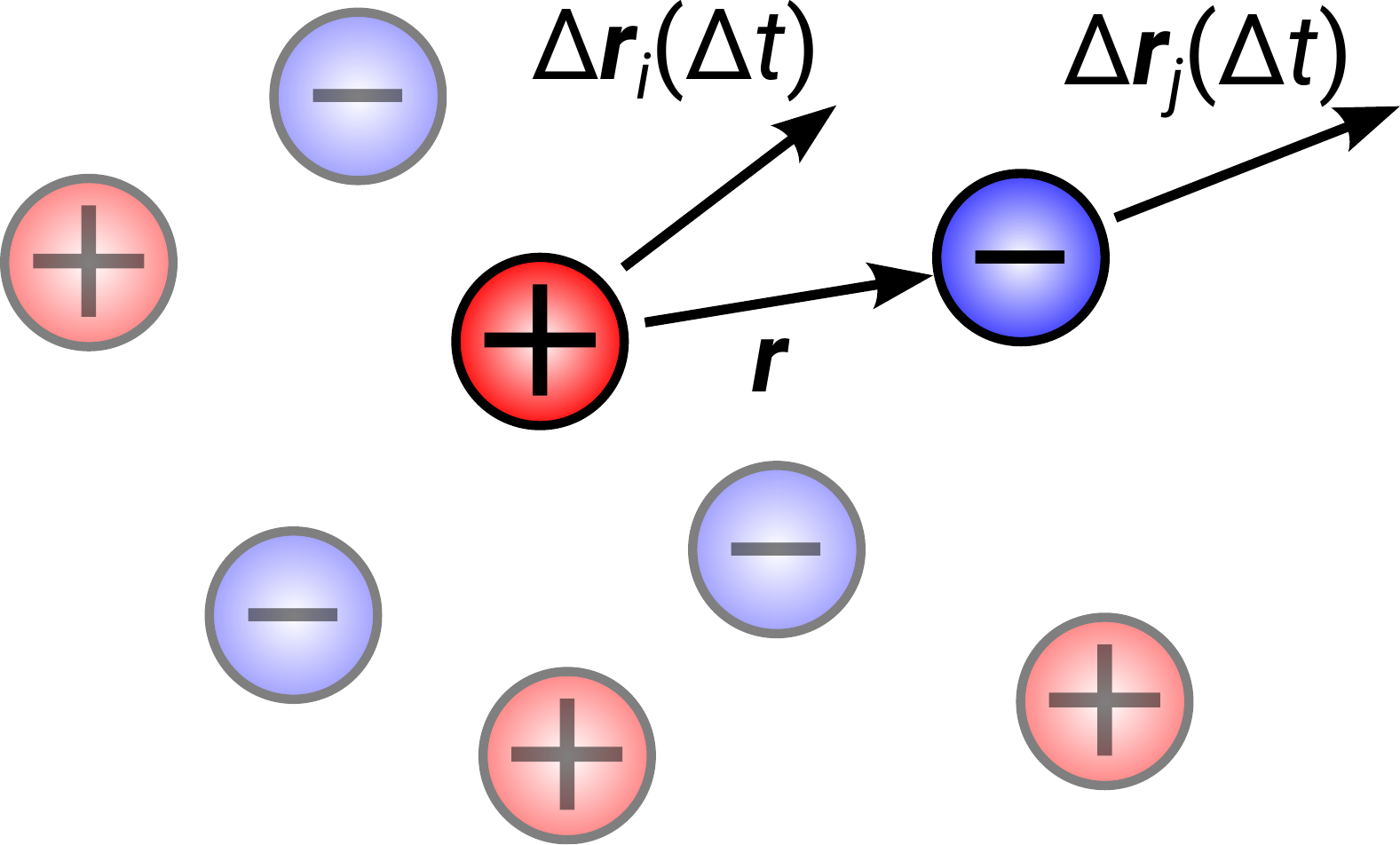}
\caption{\label{fig:scheme} Schematic illustration of the distance and displacement vectors used in this study. Motional correlations of two different ions are expressed by the average scalar product between their displacement vectors $\Delta{\bf r}_i$ and $\Delta{\bf r}_j$ in dependence of their initial distance $r$. }
\end{figure}

On the other hand, $\sigma_\pm$ and $t_\pm$ can readily be calculated from Molecular Dynamics (MD) simulations, given sufficient sampling. 
In particular, the linear-response conductivity can be extracted from equilibrium MD simulations as follows \cite{muller1995computer,wheeler2004molecular,oldiges2018understanding}: 
\begin{equation}
 \label{eq:sig}
 \sigma = \lim_{\Delta t\rightarrow \infty}\frac{e^2}{6 V\, \Delta t\, k_\mathrm{B}T}\sum_{i=1}^{N}\sum_{j=1}^{N}\,z_i z_j\,\langle\Delta{\bf r}_i\,\Delta{\bf r}_j\rangle\text{.}
\end{equation}
Here, $z_i$ and $z_j$ the valencies of ions $i$ and $j$ contained in volume $V$, $e$ the elementary charge, $k_\mathrm{B}T$ the thermal energy and $\Delta{\bf r}_i$ and $\Delta{\bf r}_j$ are the displacement vectors of ions $i$ and $j$ during lag time $\Delta t$. 
If well-defined and long-lived ion pairs of cations and anions existed, it is obvious that the net conductivity would be reduced by their presence, as such pairs would contribute to the \emph{mass} transport (and thus to the diffusion coefficients $D_\pm$), but not to the \emph{charge} transport measured by $\sigma$. 
Such a reduction is indeed seen from Eq. \ref{eq:sig} for cation-anion pairs that move cooperatively into a certain direction due to the fact that $\langle\Delta{\bf r}_i\,\Delta{\bf r}_j\rangle>0$ in this case, which dimishes $\sigma$ (see sketch in Figure \ref{fig:scheme}). 
However, the picture of distinct ion pairs is generally an oversimplification: 
First, larger ion clusters might form in an electrolyte \cite{haskins2014computational,lesch2014combined,molinari2018effect,molinari2019general,wettstein2022controlling}, and second, ion pairs or clusters are temporal in nature, i.e. they continuously disintegrate and reform \cite{zhao2009there,haskins2014computational,lesch2014combined,wettstein2022controlling}. 
Nonetheless, distinct ions move correlated (or anticorrelated) in any non-ideal electrolyte with finite concentration, which consequently affects the value of $\sigma$. 

To study these ionic correlations in an electrolyte with monovalent ions ($z_\pm = \pm 1$) in more detail, we define 
\begin{subequations}
\label{eq:sigij}
\begin{equation}
 \sigma_{++} = \lim_{\Delta t\rightarrow \infty}\frac{e^2}{6 V\, \Delta t\, k_\mathrm{B}T}\sum_{i=1}^{N_+}\sum_{j=1}^{N_+}\,\langle\Delta{\bf r}_i\,\Delta{\bf r}_j\rangle
\end{equation}
\begin{equation}
 \sigma_{+-} = \lim_{\Delta t\rightarrow \infty}\frac{-e^2}{6 V\, \Delta t\, k_\mathrm{B}T}\sum_{i=1}^{N_+}\sum_{j=1}^{N_-}\,\langle\Delta{\bf r}_i\,\Delta{\bf r}_j\rangle
\end{equation}
\begin{equation}
 \sigma_{--} = \lim_{\Delta t\rightarrow \infty}\frac{e^2}{6 V\, \Delta t\, k_\mathrm{B}T}\sum_{i=1}^{N_-}\sum_{j=1}^{N_-}\,\langle\Delta{\bf r}_i\,\Delta{\bf r}_j\rangle
\end{equation}
\end{subequations}
with $N_+$ and $N_-$ being the numbers of cations and anions, respectively, and $N_+ + N_- = N$ as well as $N_+ = N_-$ due to electroneutrality. 
In this way, Eq. \ref{eq:sig} may be decomposed into individual contributions arising from the collective motion of cations and cations, anions and anions, as well as cations and anions \cite{wohde2016li+,vargas2020dynamic} (the former two additionally containing the self-diffusion of the respective ion species): 
\begin{equation}
 \label{eq:sigterms}
 \sigma = \sigma_{++} + 2\sigma_{+-} + \sigma_{--}
\end{equation}
With the additional definitions 
\begin{subequations}
 \label{eq:sigpm}
 \begin{equation}
  \sigma_+ = \sigma_{++} + \sigma_{+-}
 \end{equation}
 and
 \begin{equation}
  \sigma_- = \sigma_{+-} + \sigma_{--}
 \end{equation}
\end{subequations}
Eq. \ref{eq:sigtot} is recovered. 

Besides these general definitions valid for any electrolyte, ILs by definition lack a neutral solvent, such that the system only consists of cations and anions. 
This has the important consequence that in periodic systems momentum conservation affects the charge transport because there are no solvent molecules that can exchange momentum with the ions. 
While this was already shown in early analytical calculations on molten salts \cite{sundheim1956transference,sundheim1964transference}, Kashyap et al. more recently confirmed the same mechanism for ILs via both analytical calculations and MD simulations \cite{kashyap2011charge}. 
A similar impact of momentum conservation on the ionic cross correlations has been observed for other quasi-binary mixtures \cite{dong2018efficient,pfeifer2021quantification}. 
In particular, because the center of mass of the system is at rest, 
\begin{equation}
 \label{eq:momconv}
 m_+\sum_{i=1}^{N_+}\Delta{\bf r}_i + m_-\sum_{i=1}^{N_-}\Delta{\bf r}_i = 0\text{,}
\end{equation}
where $m_\pm$ are the masses of cations and anions, respectively. 
Multiplying this expression by $\sum_{j=1}^{N_+}\Delta{\bf r}_j$ (or $\sum_{j=1}^{N_-}\Delta{\bf r}_j$) and 
taking the average yields 
\begin{equation}
 \label{eq:momconv2}
 m_+\sum_{i=1}^{N_+}\sum_{j=1}^{N_+}\langle\Delta{\bf r}_i\Delta{\bf r}_j\rangle + m_-\sum_{i=1}^{N_-}\sum_{j=1}^{N_+}\langle\Delta{\bf r}_i\Delta{\bf r}_j\rangle = 0\text{.}
\end{equation}
Inserting all possible combinations of the expressions in Eq. \ref{eq:sigij} into Eq. \ref{eq:momconv2}, using the short-hand notations from Eqs. \ref{eq:sigterms} and \ref{eq:sigpm}, 
accounting for the valencies $z_i$ and $z_j$ in Eq. \ref{eq:sig} and rearranging we arrive at 
\begin{subequations}
\label{eq:massratio}
\begin{equation}
 \frac{\sigma_{++}}{\sigma_{+-}} = \frac{m_-}{m_+}
\end{equation}
\begin{equation}
 \frac{\sigma_{--}}{\sigma_{+-}} = \frac{m_+}{m_-}
\end{equation}
\begin{equation}
 \frac{\sigma_{++}}{\sigma_{--}} = \frac{m_-^2}{m_+^2}
\end{equation}
\begin{equation}
 \frac{\sigma_{+}}{\sigma_{-}} = \frac{m_-}{m_+}\text{.}
\end{equation}
\end{subequations}
A particularly interesting implication pointed out by Kashyap et al. \cite{kashyap2011charge} is that $\sigma_{+-}>0$, which mathematically arises from the fact that $\sigma_{++}>0$ and $\sigma_{--}>0$ due to the dominant self-diffusion terms. 
This is in clear contrast to what is typically found for ternary electrolytes including a solvent, in which the cooperative motion of cations and anions reduces $\sigma$, that is, $\sigma_{+-}<0$. 
Because in an IL the motion of any ion has to be compensated by the motion of all other ions, on a global scale, cations and anions (but also ions with equal charges) move anticorrelated, i.e. in opposite directions \cite{kashyap2011charge}, such that $\langle\Delta{\bf r}_i\,\Delta{\bf r}_j\rangle_{+-}<0$, resulting in $\sigma_{+-}>0$ due to $z_iz_j = -1$. 

Nonetheless, one would intuitively expect that locally, neighboring ions with opposite charges move correlated. 
Via a distance-resolved analysis of the $\sigma_{XY}$ in Eq. \ref{eq:sigterms} (with $X$ and $Y$ denoting the two ion species \lq $+$\rq\ and \lq $-$\rq), Tu et al. \cite{tu2014spatial2} showed that in ILs, neighboring ions indeed move correlated, while the dominating anticorrelated motion emerges only for larger interionic separations. 
Interestingly, also equally charged ions displayed locally correlated dynamics \cite{tu2014spatial2}. 
Furthermore, Tu et al. demonstrated that qualitatively, the same features are also found for conventional aqueous electrolytes \cite{tu2014spatial,matubayasi2019spatial}, although $\sigma_{+-}<0$ as naively expected. 
In a recent paper, we could also confirm the correlated motion between lithium ions and their anionic solvation shell in IL/Li-salt mixtures with varying salt fractions \cite{wettstein2022controlling}. 

In the present contribution, we aim to understand the distance dependence of the $\sigma_{XY}$ in more detail. 
To this end, we derive an analytical framework to capture the distance dependence of $\sigma_{XY}$ based on a hydrodynamic theory that has originally been developed to calculate finite-size effects of the self-diffusion in periodic systems \cite{beenakker1986ewald,dunweg1993molecular,dunweg1993molecular2,yeh2004system,gabl2012computational}. 
These theoretical predictions are then compared to MD simulation data for a liquid carbonate electrolyte (CE) and an IL. 
The remainder of this paper is organized as follows: 
In section \ref{sec:sim}, we describe the technical aspects of the MD simulations, whereas in section \ref{sec:hydrotheory}, we develop our hydrodynamic framework for pair diffusion in periodic systems and compare it to the distance-dependent ion correlations extracted from the MD data. 
We then study the time dependence of these ionic correlations in light of our theory in section \ref{sec:timedep}. 
Finally, in section \ref{sec:conc} we conclude and give an outlook on how our framework could contribute to related topics.

\section{Simulation Details}
\label{sec:sim}

The MD simulations have been performed with the simulation code \emph{Lucretius} developed at the University of Utah 
using the APPLE\&P polarizable force field \cite{borodin2009polarizable,bedrov2019molecular}. 
For the CE, we reused MD trajectories from an earlier study \cite{oldiges2018understanding}, that is, an equimolar mixture of ethylene carbonate (EC) and dimethyl carbonate (DMC) with $1$ mol/L lithium bis(trifluoromethane)sulfonimide (LiTFSI). 
In addition, we simulated two ILs, namely 1-ethyl-3-methylimidazolium TFSI ([EMIm][TFSI]) and EMIm tetrafluoroborate ([EMIm][BF$_4$]). 
For the sake of clarity, only the data for [EMIm][TFSI] is shown in the main text, whereas the corresponding data for [EMIm][BF$_4$] is given in the Appendix. 
The ILs contained $256$ ion pairs inside a cubic simulation box. 
In addition to the systems with standard masses, comparative simulations with artificially modified masses have been performed to assess the impact of momentum conservation on the transport properties. 
To this purpose, the cation masses were scaled by a factor of $1/\sqrt{2}$, while the anion masses have been increased by a factor $\sqrt{2}$. 

The systems were equilibrated for $5$ ns in the $NpT$ ensemble, followed by subsequent production runs of $100$ ns in the $NVT$ ensemble at $298$ K, resulting in box lengths of $47.7589$ \AA\ and $40.5430$ \AA\ for [EMIm][TFSI] and [EMIm][BF$_4$], respectively. 
Both the temperature and the pressure of the system were maintained by a Nos\'{e}-Hoover chain thermostat (coupling frequency $0.01$ fs$^{-1}$) and barostat (coupling frequency $0.0005$ fs$^{-1}$) \cite{martyna1992nose}, while periodic boundary conditions were applied in all dimensions. 
Electrostatic interactions have been treated by the Ewald summation technique with a cut-off radius of $12$ \AA, an inverse Gaussian charge width of $0.23$ \AA$^{-1}$, and $7\times 7\times 7$ vectors for the reciprocal space. 
Lennard-Jones interactions have been truncated at $12$ \AA, beyond which a continuum-model dispersion correction was applied. 
All bonds were constrained by the SHAKE algorithm \cite{ryckaert1977numerical,palmer1993direct}. 
A multiple-time-step integration scheme \cite{martyna1994constant,martyna1996explicit} was used to integrate the equations of motion, where a time step of $0.5$ fs has been used for bonds and angles. 
For torsions and non-bonded interactions up to a distance of $7$ \AA, a time step of $1.5$ fs was used, and finally, for non-bonded interactions between atoms separated more than $7$ \AA\ and the reciprocal part of the Ewald summation, a time step of $3$ fs was used. 
The induced dipoles were determined iteratively where the corresponding dipole-dipole interactions were scaled to zero by a tapering function between $11$ and $12$ \AA. 
The pressure tensor was dumped every $0.9$ ps to calculate the viscosity (section \ref{ssec:decay} and Appendix \ref{sec:visc}). 

The CE from the previous study \cite{oldiges2018understanding} was simulated in an $NpT$ ensemble, for which the unwrapping the ions' coordinates from the primary simulation box into real space can be problematic \cite{von2020systematic,kulke2022reversible}. 
In the present work, we observed a similar effect for the pair diffusion. 
Therefore, the algorithm described in Ref. \citenum{von2020systematic} was employed.

\section{Hydrodynamic Theory}
\label{sec:hydrotheory}

\subsection{Distance Dependence of Ionic Correlations}
\label{ssec:theory}

We start by characterizing the distance dependence of the ionic correlations $\langle\Delta{\bf r}_i\,\Delta{\bf r}_j\rangle$ in Eqs. \ref{eq:sig} and \ref{eq:sigij}. 
To this end, we define the dynamical correlation between two distinct ions as a function of their initial separation $r$: 
\begin{equation}
 \label{eq:rhodef}
 \rho_{XY}(r) = \langle\Delta{\bf r}_i\,\Delta{\bf r}_j \vert r_{ij}(0) = r \rangle_{XY}
\end{equation}
where $X$ and $Y$ denote the ion species as above. 
The index at the $\langle\ldots\rangle_{XY}$ bracket indicates that the average is taken for a given pair type. 
In addition, only ion pairs with a given initial separation $r_{ij}(0) = r$ are averaged, as indicated by the conditional expression in Eq. \ref{eq:rhodef}. 

\begin{figure*}
 \subfloat[\label{fig:rhor_cb_30ps}]{
 \includegraphics[width=0.5\textwidth]{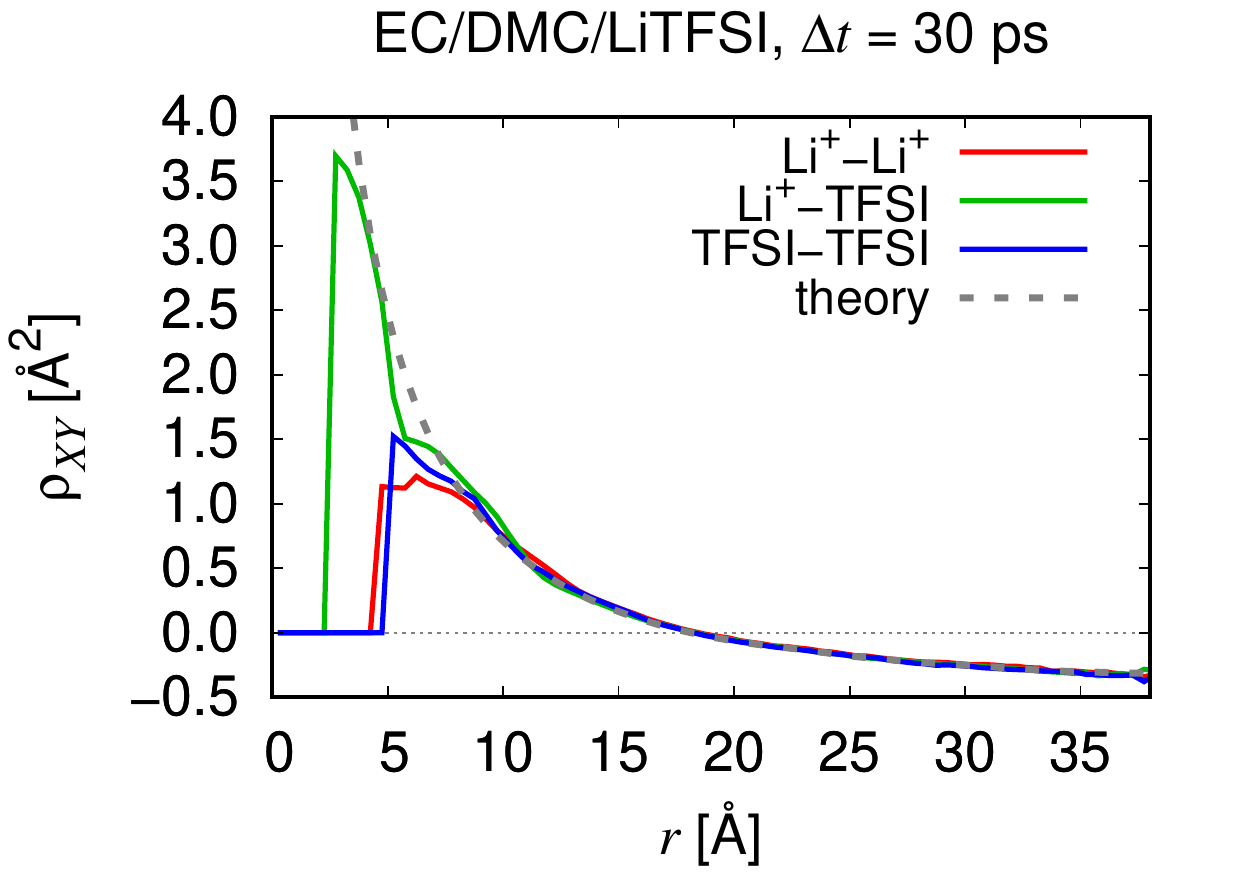}}
 \subfloat[\label{fig:rhor_et_100ps}]{
 \includegraphics[width=0.5\textwidth]{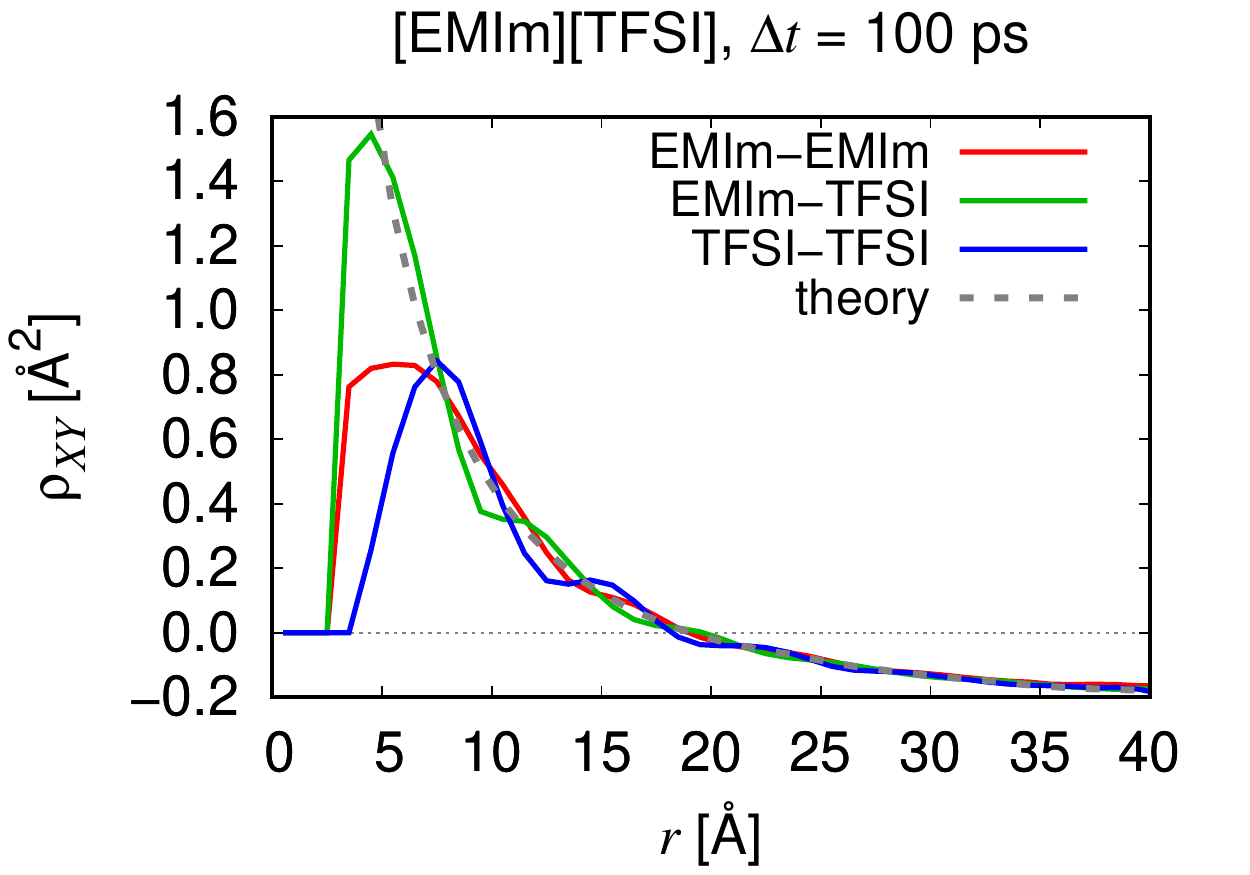}}
\\
 \subfloat[\label{fig:drhor_cb_30ps}]{
 \includegraphics[width=0.5\textwidth]{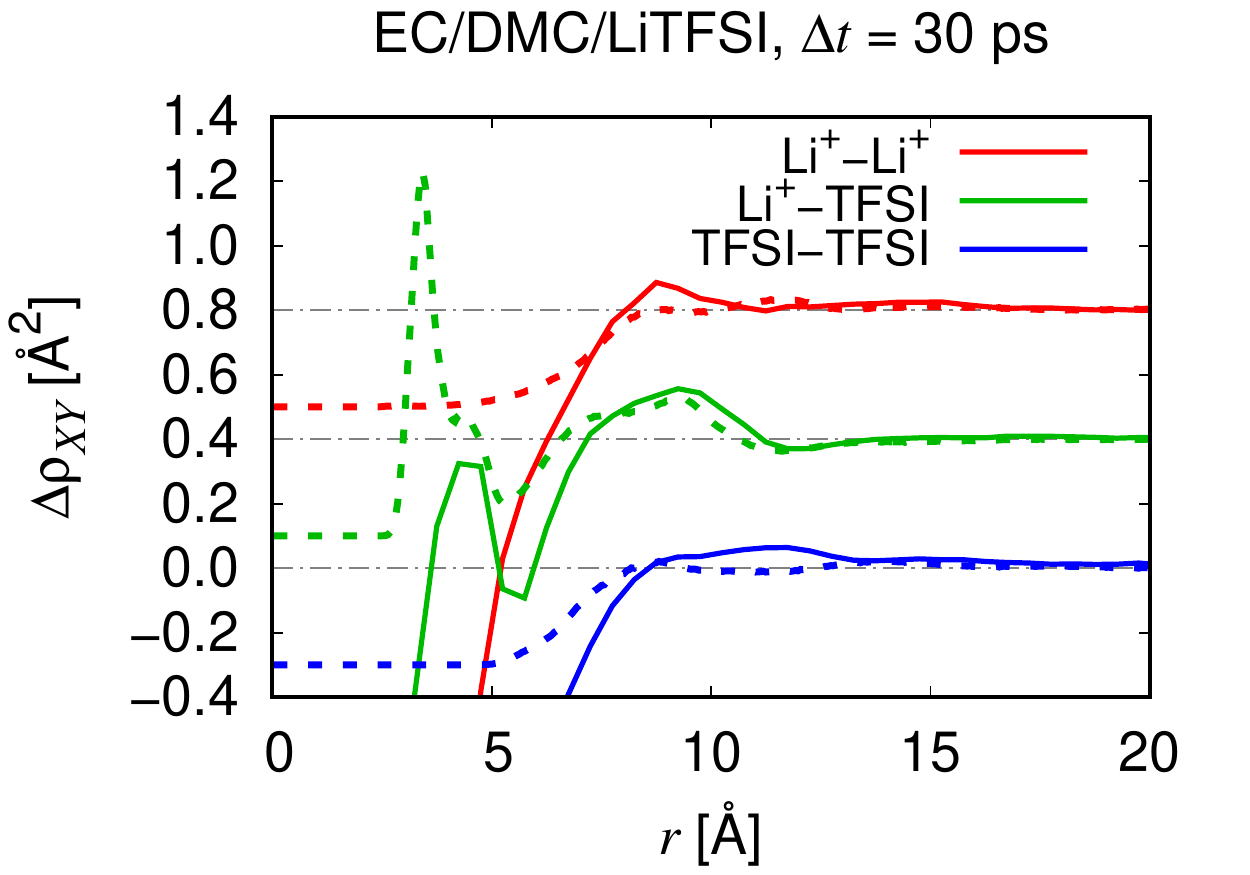}}
 \subfloat[\label{fig:drhor_et_100ps}]{
 \includegraphics[width=0.5\textwidth]{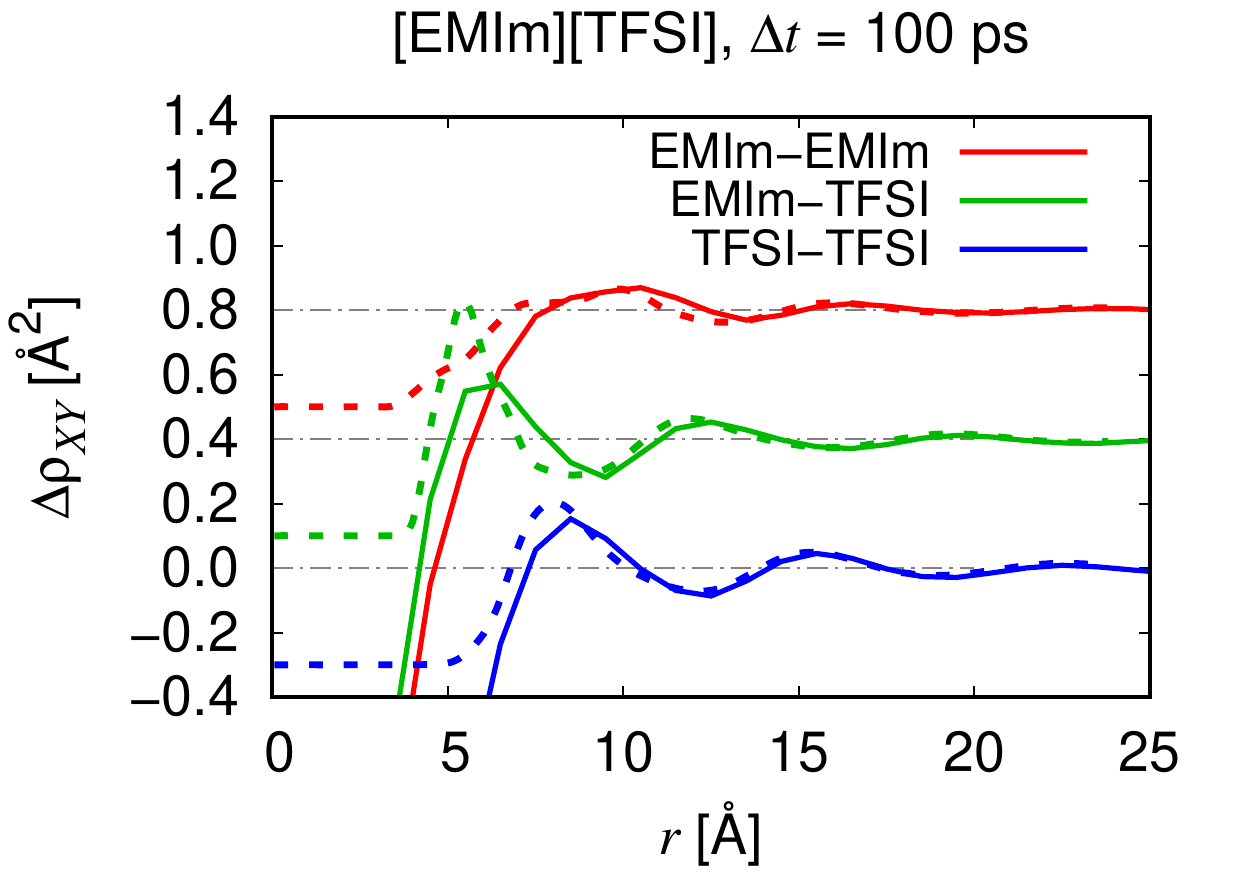}}
\caption{\label{fig:rhor} (a), (b): Distance-dependent correlation $\rho_{XY}(r) = \langle\Delta{\bf r}_i\,\Delta{\bf r}_j \vert r_{ij}(0) = r \rangle_{XY}$ between ion pair types $X$ and $Y$ in dependence of their initial distance $r$. (c), (d): Deviations of the simulation data from the hydrodynamic fit ($\Delta\rho_{XY} = \rho_{XY} - \rho_\mathrm{PBC}$, solid curves) in comparison with the scaled and shifted RDFs (dashed curves). Several curves in (c) and (d) have been shifted for better visibility. The uncertainties are in the range of the line thickness.}
\end{figure*}

The individual $\rho_{XY}$ are shown in Figure \ref{fig:rhor_cb_30ps} for the CE with $\Delta t = 30$ ps and Figure \ref{fig:rhor_et_100ps} for the IL with $\Delta t = 100$ ps. 
Here, the $\Delta t$-values have been chosen such that the dynamics is still subdiffusive to avoid that the distance between the ions changes too much during $\Delta t$, which would blur the $\rho_{XY}$-curves in Figure \ref{fig:rhor}. 
However, the behavior for larger $\Delta t$ up to a few nanoseconds is qualitatively the same, which we analyze below in section \ref{sec:timedep}. 
From Figure \ref{fig:rhor} we observe that the curves of all pair types display a peak with $\rho_{XY}>0$ at short initial separations (i.e. at about $5-10$ \AA), demonstrating that locally, all ion pairs move correlated. 
Similar observations have been made previously by Tu et al. \cite{tu2014spatial,tu2014spatial2}. 
The short-range peak for the cation-anion correlation is larger than the respective peaks of $\rho_{++}$ and $\rho_{--}$ and shifted to shorter distances. 
This observation can be rationalized by the local ordering, i.e., the nearest-neighbor shell of a cation is essentially composed of anions and vice versa \cite{canongia2006nanostructural,hardacre2007structure,zhao2009there,oldiges2018understanding}, resulting in peak positions of $\rho_{+-}$ shifted to shorter $r$ and showing a larger magnitude. 
In case of the IL, all $\rho_{XY}$ show a decay superimposed with minor oscillations at intermediate distances, presumably related to its long-ranged ordering \cite{canongia2006nanostructural,hardacre2007structure,zhao2009there}. 
At large separations, all $\rho_{XY}$ become increasingly negative for both systems, although the curves vary only slowly with $r$, demonstrating that the ions move anticorrelated, which is in agreement with the findings of Tu et al. \cite{tu2014spatial,tu2014spatial2}. 

In previous works \cite{kashyap2011charge,tu2014spatial2}, the anticorrelated motion has (at least partly) been attributed to the momentum conservation constraint. 
That is, because locally all ion types move correlated, momentum conservation can only be realized by a compensating counterflux of ions at larger length scales. 
However, from Figure \ref{fig:rhor_cb_30ps} we observe anticorrelated motion at large $r$ also for the CE, in line with previous results for an aqueous ternary electrolyte \cite{kashyap2011charge}. 
In fact, the qualitative shape of all $\rho_{XY}$ in Figure \ref{fig:rhor} is remarkably similar for larger $r$, pointing towards a universal feature. 
Empirically, we find that the decay of $\rho_{XY}$ scales as $1/r$, which is indicative of hydrodynamic interactions \cite{doi1988theory,dunweg1993molecular,dunweg1993molecular2,yeh2004system}. 
As the $1/r$-scaling is theoretically expected even when neglecting inertial forces \cite{doi1988theory}, the incompressibility of the medium must play an important role, too. 
Indeed, early analytical descriptions of the ionic conductivity assumed the existence of a hydrodynamic flow field around a given central ion \cite{fuoss1963conductance,lee1978conductance,ebeling1979electrolytic,altenberger1983theory}. 
Furthermore, the hydrodynamic picture is also in line with our recent findings for IL/Li-salt blends, where we observed that a given anion still moves cooperatively with a nearby lithium ion even after disenganging from its coordination shell \cite{wettstein2022controlling}. 
In the following, we therefore provide a theoretical basis for the observations in Figure \ref{fig:rhor_cb_30ps} and \ref{fig:rhor_et_100ps}.

\subsection{Pair Diffusion in Periodic Systems}
\label{ssec:theory2}

\subsubsection{Two Derivations of the Pair Diffusion Tensor}
\label{sssec:dtens}

Starting from Eq. \ref{eq:sig}, we define the diffusion tensor with the elements \cite{dunweg1993molecular,gabl2012computational}
\begin{equation}
  \label{eq:dcross}
  D_{ij}^{\beta\gamma} = \lim_{\Delta t\rightarrow\infty}\frac{\langle \Delta r_i^\beta\,\Delta r_j^\gamma\rangle}{6\Delta t}
\end{equation}
describing the pair diffusion of ions $i$ and $j$, where $\beta$ and $\gamma$ denote the spatial compounds of $\Delta{\bf r}_i$ and $\Delta{\bf r}_j$, respectively. 
For $i=j$ in isotropic systems, the trace of this expression reduces to the well-known Einstein relation: 
\begin{equation}
  \label{eq:dself}
  D_i = \lim_{\Delta t\rightarrow\infty}\frac{\langle\Delta{\bf r}_i^2\rangle}{6\Delta t}
\end{equation}
Rather than focusing on explicit pairs of ions, one may define the diffusion tensor as a function of the interionic separation ${\bf r}={\bf r}_j-{\bf r}_i$ for $i\neq j$ \cite{dunweg1993molecular,gabl2012computational}, similar in spirit to Eq. \ref{eq:rhodef}: 
\begin{equation}
 \label{eq:dtensr}
  {\bf D}_{ij} = {\bf D}({\bf r}_j-{\bf r}_i)
\end{equation}
For an infinite system, the diffusion tensor is given to first order (i.e. within the approximation of point particles valid for large $r$) by the so-called Oseen tensor \cite{doi1988theory}
\begin{equation}
  \label{eq:treal}
  {\bf D}_\infty({\bf r}) = \frac{k_\mathrm{B}T}{8\pi\eta r}(1+\hat{\bf r}\hat{\bf r})
\end{equation}
that describes the hydrodynamic interactions in a fluid with viscosity $\eta$. 
For a cubic periodic system with box length $L$, analogous expressions were derived \cite{dunweg1993molecular,yeh2004system}, which we sketch in the following.

\paragraph{Green-Kubo relation}

D\"unweg \cite{dunweg1993molecular} expressed Eq. \ref{eq:dtensr} via the Green-Kubo relation 
\begin{equation}
  \label{eq:dgk}
  {\mathbf D}_{ij}({\mathbf r}) = \int_0^\infty\,dt\,\langle{\mathbf u}(0,0){\mathbf u}({\mathbf r},t)\rangle\text{,}
\end{equation}
where ${\mathbf u}({\mathbf r},t)$ is the flow field in a continuous representation. 
From a set of discrete particles with positions $\lbrace{\mathbf r}_i\rbrace$, the latter can be obtained via ${\mathbf u}({\mathbf r},t) = (V/N)\sum_{i=1}^N\,{\mathbf v}_i\delta({\mathbf r}-{\mathbf r}_i)$, with $\lbrace{\mathbf v}_i\rbrace$ being the particles' velocities. 
Due to periodicity, ${\mathbf u}({\mathbf r},t)$ can be expressed by its Fourier modes, i.e. 
 \begin{align}
  \begin{split}
  \label{eq:umodes}
  {\mathbf u}({\mathbf r},t) &= \sum_{\mathbf k}\tilde{{\mathbf u}}_{\mathbf k}(t)\exp{(i{\mathbf k}\cdot{\mathbf r})} \\
  \tilde{{\mathbf u}}_{\mathbf k}(t) &= \frac{1}{N}\sum_{i=1}^N {\mathbf v}_i(t)\exp{(-i{\mathbf k}\cdot{\mathbf r}_i)}\text{,}
  \end{split}
 \end{align}
where ${\mathbf k} = 2\pi{\mathbf n}/L$ is a reciprocal lattice vector with ${\mathbf n} = (n_x n_y n_z)^T$ and $n_x$, $n_y$, $n_z \in \mathbb{Z}$. 
The diffusion tensor in the case of uncorrelated modes then reads \cite{dunweg1993molecular}
\begin{equation}
 \label{eq:tfourierflowfield}
 {\mathbf D}_\mathrm{PBC}({\mathbf r}) = \sum_{{\mathbf k} \neq {\mathbf 0}}(1-\hat{\mathbf k}\hat{\mathbf k})\exp{(i{\mathbf k}\cdot{\mathbf r})} \int_0^\infty dt\,\langle\tilde{{\mathbf u}}_{\mathbf k}(0)\tilde{{\mathbf u}}_{\mathbf k}(t)\rangle
\end{equation}
(note that $(1-\hat{\mathbf k}\hat{\mathbf k})$ projects on the transversal modes relevant for an incompressible system). 
The zeroth mode is excluded, as it describes to the net motion of the entire system \cite{dunweg1993molecular,yeh2004system}. 
D\"unweg evaluated the Green-Kubo integral on the right-hand side of Eq. \ref{eq:tfourierflowfield} via the Mori-Zwanzig formalism, yielding \cite{dunweg1993molecular,gabl2012computational}
\begin{equation}
 \label{eq:morizwanzig}
 \int_0^\infty dt\,\langle\tilde{{\mathbf u}}_{\mathbf k}(0)\tilde{{\mathbf u}}_{\mathbf k}(t)\rangle = \frac{(k_\mathrm{B}T)^2}{k^2 V^2 \int_0^\infty dt\,\langle P_{\beta\gamma}(0)P_{\beta\gamma}(t)\rangle}\text{,}
\end{equation}
where the 
\begin{equation}
 \label{eq:ptens}
 P_{\beta\gamma}(t) = \frac{1}{V}\sum_{i=1}^{M}m_i v_{i\beta}(t) v_{i\gamma}(t) + F_{i\beta}(t) r_{i\gamma}(t)
\end{equation}
are the off-diagonal elements of the pressure tensor ($\beta\neq\gamma$), $m_i$ is the mass, $v_{i\beta}$ the velocity, $F_{i\beta}$ the force and $r_{i\beta}$ the position of particle $i$ in $\beta$-direction, respectively. 
Note that $M$ in Eq. \ref{eq:ptens} denotes the total number of particles/atoms as opposed to the number of ions $N$ and hence may also include a solvent. 
The integral in the denominator of Eq. \ref{eq:morizwanzig} is nothing else than the Green-Kubo relation for the viscosity \cite{holian1983shear,yeh2004system}
\begin{equation}
 \label{eq:gketa}
 \eta = \frac{V}{k_\mathrm{B}T} \int_0^\infty dt \langle P_{\beta\gamma}(0)P_{\beta\gamma}(t)\rangle\text{,}
\end{equation}
such that Eq. \ref{eq:tfourierflowfield} becomes 
\begin{equation}
  \label{eq:tfourier}
  {\bf D}_\mathrm{PBC}({\bf r}) = \frac{k_\mathrm{B}T}{V\eta}\sum_{{\bf k} \neq {\bf 0}}\frac{(1-\hat{\bf k}\hat{\bf k})\exp{(i{\bf k}\cdot{\bf r})}}{k^2}\text{.}
\end{equation}

\paragraph{Stokes equation}

An alternative route to Eq. \ref{eq:tfourier} is via the Stokes equation 
\begin{equation}
 \label{eq:stokes}
 \eta\nabla^2{\mathbf u}({\mathbf r}) = \nabla p({\mathbf r})-\left(\delta({\mathbf r})-\frac{1}{V}\right){\mathbf F}
\end{equation}
as shown by Yeh and Hummer \cite{yeh2004system}, where $p$ is the pressure and ${\mathbf F}$ a perturbative force acting on a point-like particle. 
The term $1/V$ ensures that the net force acting on the periodic cell is zero, leading to the exclusion of the zeroth mode (see above). 
For an incompressible fluid with $\nabla{\mathbf u}=0$ the divergence of Eq. \ref{eq:stokes} simplifies to 
\begin{equation}
 \label{eq:divstokes}
 \nabla^2 p({\mathbf r}) = {\mathbf F}\cdot\nabla\left(\delta({\mathbf r})-\frac{1}{V}\right)\text{.} 
\end{equation}
Eqs. \ref{eq:stokes} and \ref{eq:divstokes} can be transformed into Fourier space via Eq. \ref{eq:umodes} and $p({\mathbf r},t) = \sum_{\mathbf k}\tilde{p}_{\mathbf k}(t)\exp{(i{\mathbf k}\cdot{\mathbf r})}$, giving 
\begin{equation}
  -\eta k^2\tilde{{\mathbf u}}_{\mathbf k} = i{\mathbf k}\tilde{p}_{\mathbf k} - (1+\delta_{\mathbf k}){\mathbf F}
\end{equation}
and 
\begin{equation}
  -k^2\tilde{p}_{\mathbf k} = i(1-\delta_{\mathbf k}){\mathbf k}\cdot{\mathbf F}\text{.} 
\end{equation}
Eliminating $\tilde{p}_{\mathbf k}$, rearranging, summing over all modes and using ${\mathbf u} = (k_\mathrm{B}T)^{-1}\,{\mathbf D}_\mathrm{PBC}\cdot{\mathbf F}$ results in Eq. \ref{eq:tfourier} as well.

\subsubsection{Trace of the Diffusion Tensor}

Assuming that the relative orientation of the displacement vectors $\Delta{\bf r}_i$ and $\Delta{\bf r}_j$ is statistically independent from that of ${\bf r}$, we may take the trace of the tensors in Eqs. \ref{eq:treal} and \ref{eq:tfourier} to yield 
\begin{equation}
  \label{eq:dreal}
  D_\infty(r) = \frac{k_\mathrm{B}T}{6\pi\eta r}
\end{equation}
for infinite systems and 
\begin{equation}
  \label{eq:dfourier}
  D_\mathrm{PBC}(r) = \frac{k_\mathrm{B}T}{6\pi\eta}\frac{1}{V}\sum_{{\bf k} \neq {\bf 0}}\frac{4\pi}{k^2}\exp{(i{\bf k}\cdot{\bf r})}
\end{equation}
for periodic systems. 
Thus, for an infinite system, we recover the Oseen-like decay \cite{dunweg1993molecular,dunweg1993molecular2,yeh2004system} proportional to $1/r$, whereas in a periodic system a more intricate distance dependence is found.

\subsubsection{Ewald Summation}

Due to the fact that the summation in Eq. \ref{eq:dfourier} is ill-convergent, one usually applies the Ewald summation technique  \cite{hummer1998molecular,dunweg1993molecular2,yeh2004system,gabl2012computational}, in which an additional convergence factor $\exp{(-k^2/(4\alpha^2))}$ and a short-ranged compensating real-space summation is introduced to the summands in Eq. \ref{eq:dfourier}, resulting in \cite{hummer1998molecular,yeh2004system} 
\begin{widetext}
\begin{equation}
  \label{eq:dconv}
  D_\mathrm{PBC}(r) = \frac{k_\mathrm{B}T}{6\pi\eta}\left[ \frac{1}{V}\sum_{{\bf k} \neq {\bf 0}}\frac{4\pi}{k^2}\exp{(i{\bf k}\cdot{\bf r})}\exp{\left(-\frac{k^2}{4\alpha^2}\right)} + \sum_{{\bf n}}\frac{\erfc{\left(\alpha|{\bf r} + {\bf n}L|\right)}}{|{\bf r} + {\bf n}L|} - \frac{\pi}{V\alpha^2} \right]\text{.}
\end{equation}
\end{widetext}
The finite-size correction $\Delta D_\mathrm{FSC}(r) = D_\mathrm{PBC}(r) - D_\infty(r)$ for the comparison between periodic and the infinite systems can then be written as 
\begin{widetext}
\begin{equation}
  \label{eq:rfsc}
  \Delta D_\mathrm{FSC}(r) = D_\mathrm{PBC} - D_\infty = \frac{k_\mathrm{B}T}{6\pi\eta}\left[ \frac{1}{V}\sum_{{\bf k} \neq {\bf 0}}\frac{4\pi}{k^2}\exp{(i{\bf k}\cdot{\bf r})}\exp{\left(-\frac{k^2}{4\alpha^2}\right)} + \sum_{{\bf n}\neq {\bf 0}}\frac{\erfc{\left(\alpha|{\bf r} + {\bf n}L|\right)}}{|{\bf r} + {\bf n}L|} - \frac{\erf{(\alpha\,r)}}{r} - \frac{\pi}{V\alpha^2} \right]
\end{equation}
\end{widetext}
which may be numerically evaluated. 
In the limit $r\rightarrow 0$ we recover the expression 
\begin{equation}
  \label{eq:dfsc}
  \Delta D_\mathrm{FSC}(r\rightarrow 0) = -\frac{k_\mathrm{B}T}{6\pi\eta}\frac{\xi(r\rightarrow 0)}{L}
\end{equation}
as already derived by D\"unweg \cite{dunweg1993molecular2} and Yeh and Hummer \cite{yeh2004system}, where $\xi(r\rightarrow 0) \approx 2.837297$ is a constant. 
Eq. \ref{eq:dfsc} is frequently used to calculate the finite-size correction for diffusion coefficients obtained from MD simulation data. 
Similar expressions have been derived for non-cubic box geometries \cite{hasimoto1959periodic,cao2019correction}. 

Eq. \ref{eq:rfsc} may also be converted into its dimensionless form via the dimensionless distance vector ${\bf r}/L$ and dimensionless convergence parameter $\alpha L$, 
\begin{widetext}
 \begin{align}
  \begin{split}
  \label{eq:rfscnondim}
  \Delta D_\mathrm{FSC}(r/L) &= \frac{k_\mathrm{B}T}{6\pi\eta L}\left[ \sum_{{\bf n} \neq {\bf 0}}\frac{1}{\pi n^2}\exp{\left(2 \pi i\,{\bf n}\cdot({\bf r}/L)\right)}\exp{\left(-\frac{\pi^2 n^2}{(\alpha L)^2}\right)} \right. \\
   &+ \left.\sum_{{\bf n}\neq {\bf 0}}\frac{\erfc{\left((\alpha L)|({\bf r}/L) + {\bf n}|\right)}}{|({\bf r}/L) + {\bf n}|} - \frac{\erf{\left((\alpha L)(r/L)\right)}}{r/L} - \frac{\pi^2}{(\alpha L)^2} \right] \\
   &= -\frac{k_\mathrm{B}T}{6\pi\eta}\frac{\xi(r/L)}{L}\text{,}
   \end{split}
 \end{align}
\end{widetext}
from which the overall scaling proportional to $L^{-1}$ becomes apparent. 

To calculate pair diffusion coefficients expected for the individual terms in Eq. \ref{eq:sig} in a periodic MD system, we use $D_\mathrm{PBC} = D_\infty + \Delta D_\mathrm{FSC}$ for finite $r$, for which Eq. \ref{eq:rfscnondim} yields 
\begin{equation}
 D_\mathrm{PBC}(r/L) = \frac{k_\mathrm{B}T}{6\pi\eta L}\left[\frac{L}{r} - \xi(r/L)\right]\text{.}
\end{equation}
Finally, Eq. \ref{eq:dcross} can be used to convert $D_\mathrm{PBC}(r)$ to $\rho_\mathrm{PBC}(r)$ as defined in Eq. \ref{eq:rhodef}: 
\begin{equation}
 \label{eq:fsesh}
 \rho_\mathrm{PBC}(r/L) = \frac{k_\mathrm{B}T\Delta t}{\pi\eta L}\left[\frac{L}{r} - \xi(r/L)\right]
\end{equation}
The numerically evaluated curve for $\xi(r/L)$ is shown in Figure \ref{fig:xir} in Appendix \ref{sec:ewald}. 
We observe that $\xi(r/L)$ slightly decays from its initial value to $\xi\approx 2$ for the largest possible distance in the box, i.e. $r/L = \sqrt{3}/2$. 
Note that diffusive dynamics has implicitly been assumed in Eq. \ref{eq:fsesh}, which we discuss further in section \ref{sec:timedep}.

\subsection{Flow Field of the Pair Diffusion Tensor}
\label{ssec:difftensor}

\begin{figure*}
 \subfloat[\label{fig:flow100}]{
 \includegraphics[width=0.5\textwidth]{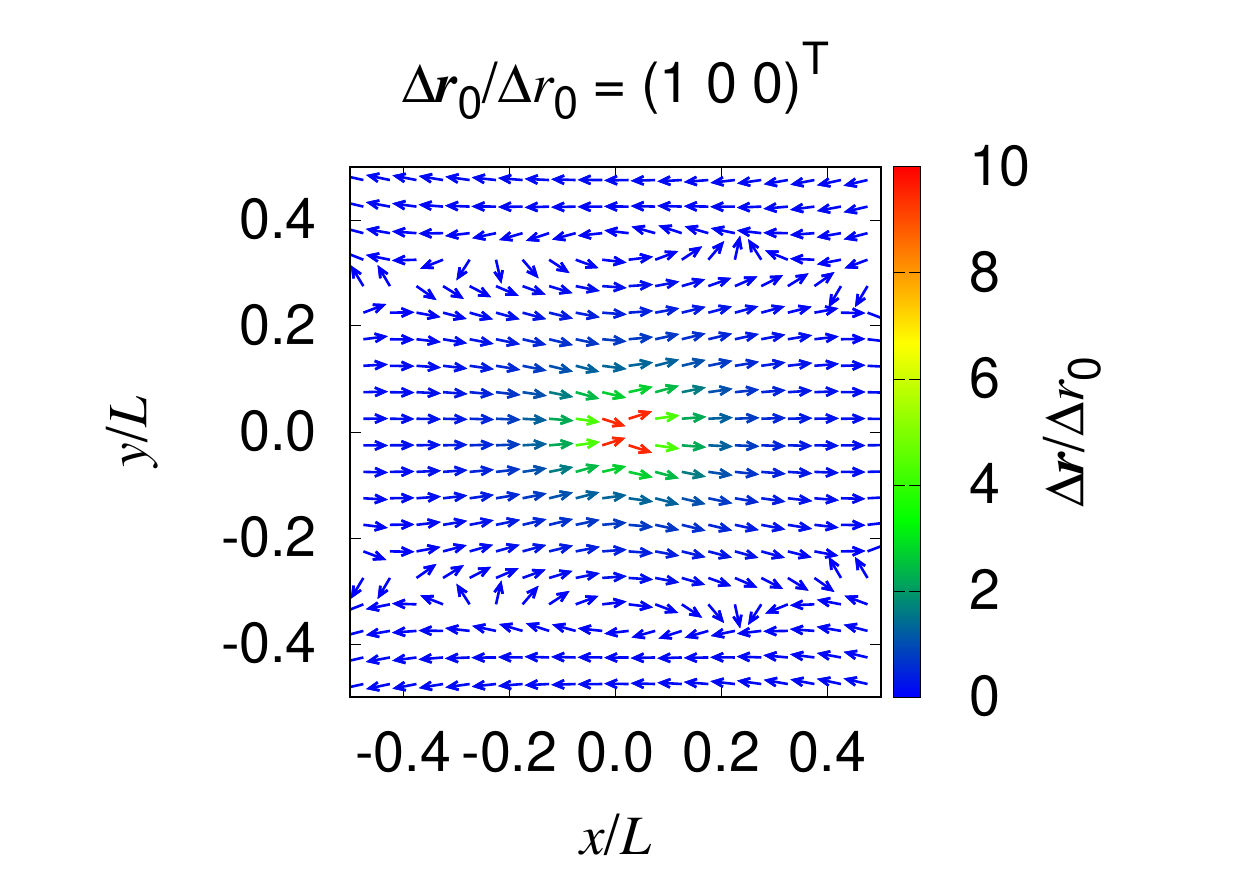}}
 \subfloat[\label{fig:flow110}]{
 \includegraphics[width=0.5\textwidth]{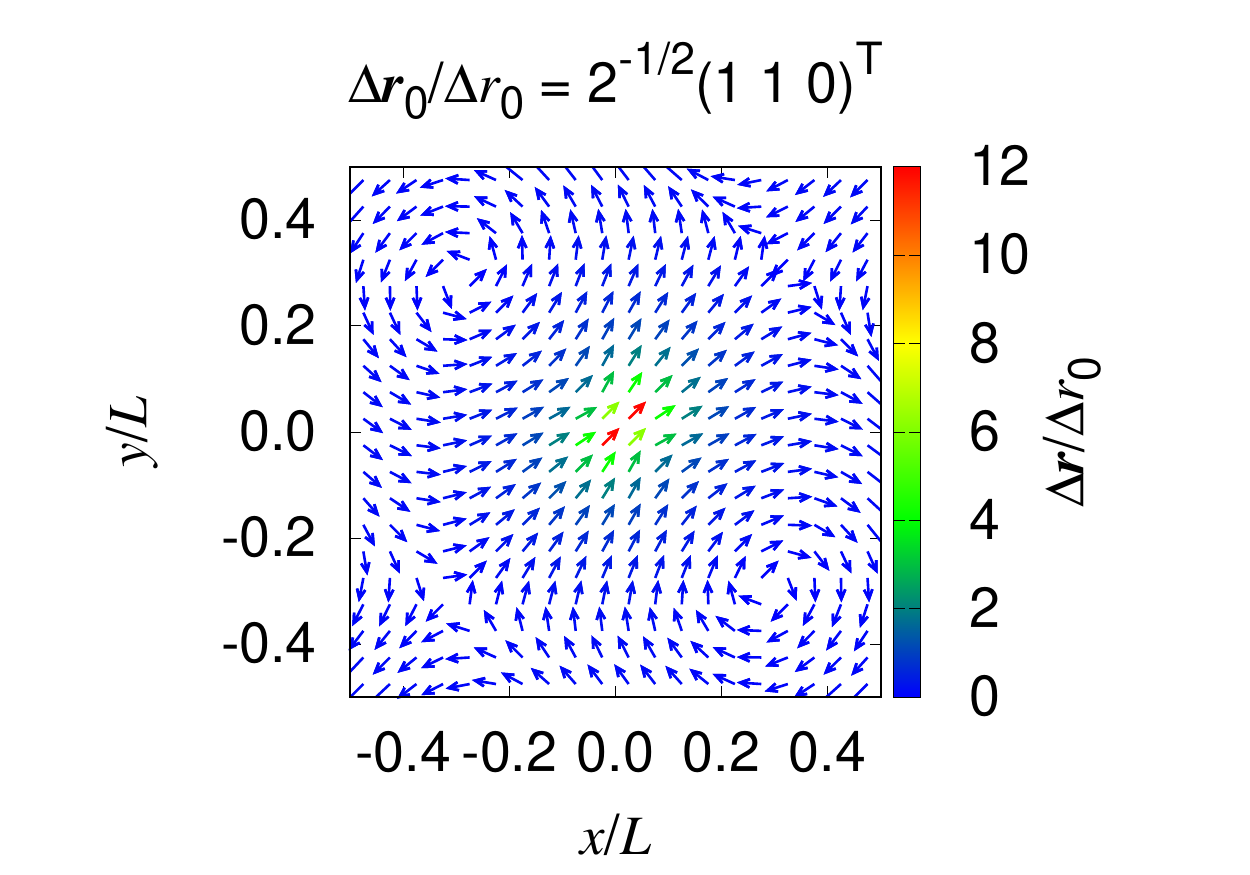}}
\\
 \subfloat[ \label{fig:flow101}]{
 \includegraphics[width=0.5\textwidth]{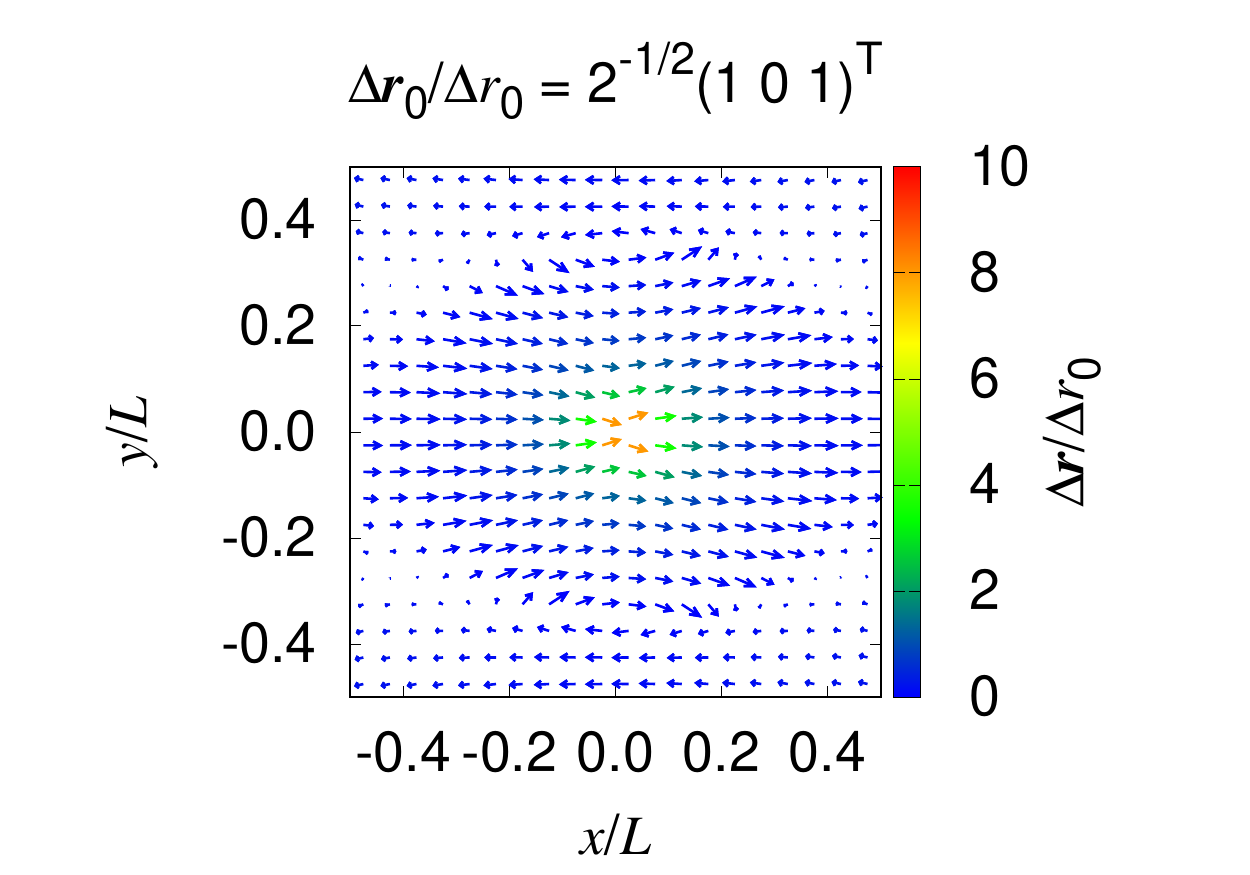}}
 \subfloat[\label{fig:flow111}]{
 \includegraphics[width=0.5\textwidth]{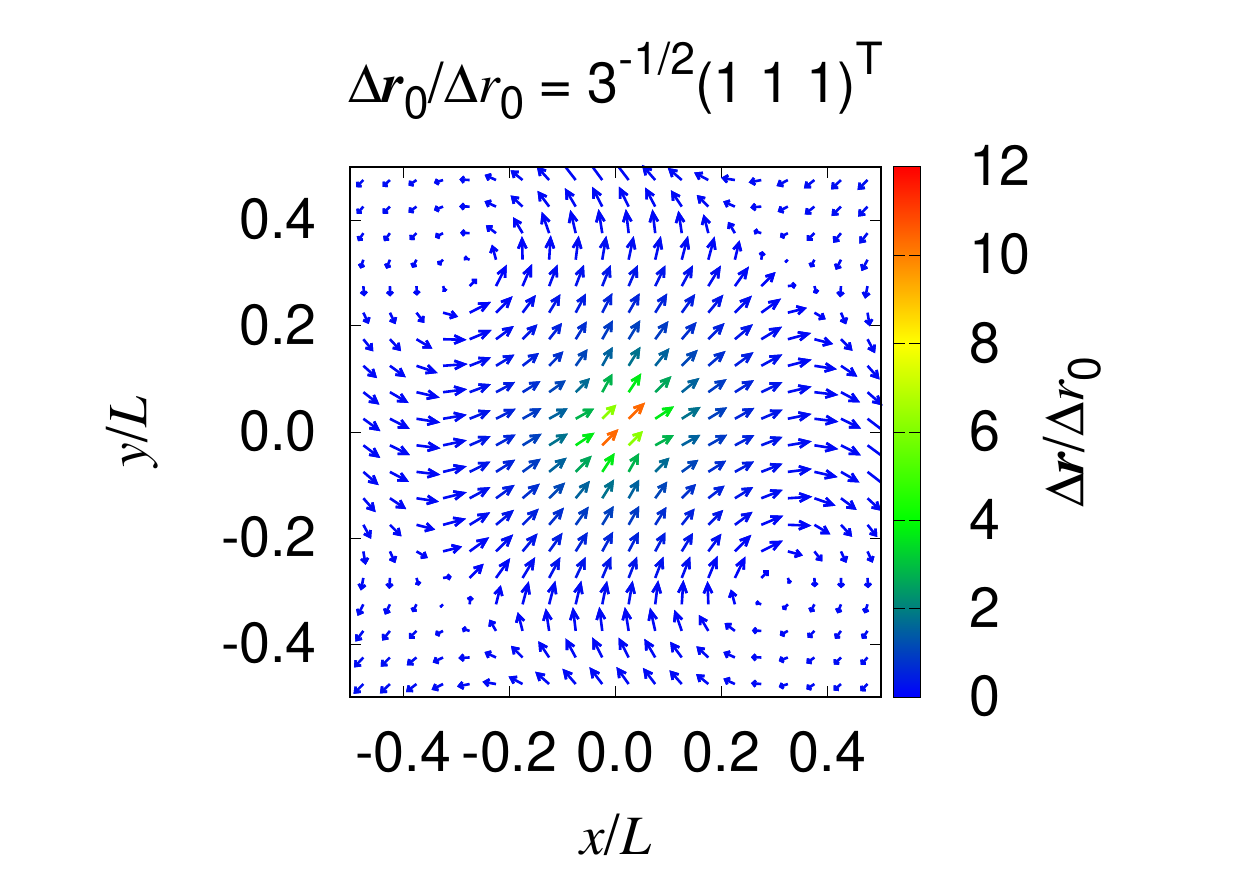}}
\caption{\label{fig:dtens} Normalized flow field generated by the diffusion tensor in Eq. \ref{eq:tfourier} projected onto the $x$,$y$-plane for different displacement vectors $\Delta{\mathbf r}_0$ of a point particle in the center of the box. The orientation of the resulting $\Delta{\mathbf r}$ vector field is shown as arrows, its relative magnitude marked by the color coding.}
\end{figure*}

Rather than taking the trace in Eqs. \ref{eq:dreal} and \ref{eq:dfourier}, we evaluated the flow field generated by the diffusion tensor in Eq. \ref{eq:tfourier} in a first step. 
Due to the ill-converging behavior of the expression in Eq. \ref{eq:tfourier}, we used the Ewald summation in analogy to the derivation of Eq. \ref{eq:dconv}, but retaining the orientational dependence. 
To this end, the real-space summands (i.e. the second, third and fourth term on the right-hand side of Eq. \ref{eq:rfsc} as well as the expression from Eq. \ref{eq:treal}) were weighted by the tensor product $(\Delta\hat{\mathbf r}_0+(\hat{\mathbf r}\cdot\Delta\hat{\mathbf r}_0)\hat{\mathbf r})$, whereas the Fourier term (first term on the right-hand side of Eq. \ref{eq:rfsc}) was scaled by $(\Delta\hat{\mathbf r}_0-(\hat{\mathbf k}\cdot\Delta\hat{\mathbf r}_0)\hat{\mathbf k})$ (cf. tensorial products in Eqs. \ref{eq:treal} and \ref{eq:tfourier}; see Appendix \ref{sec:ewald} for details). 
Here, $\Delta{\mathbf r}_0$ is the displacement vector of a point particle in the center of the box and the hats denote unit vectors. 
Unit values and unit vectors have been employed for the numerical evaluation of the flow field. 

Figure \ref{fig:dtens} shows this normalized flow field in the $x$,$y$-plane for different $\Delta{\mathbf r}_0$. 
As already indicated by Figure \ref{fig:rhor}, we observe a locally aligned flow and a global counterflux in Figures \ref{fig:flow100} and \ref{fig:flow101} when $\Delta{\mathbf r}_0$ is oriented parallel to the $x$-axis. 
Interestingly, when $\Delta{\mathbf r}_0$ acts along the diagonal, vortices appear at large distances from the center in Figures \ref{fig:flow110} and \ref{fig:flow111}. 
Qualitatively, the flow field in Figures \ref{fig:flow101} and \ref{fig:flow111} is similar than for the case when $\Delta{\mathbf r}_0$ lies in the $x$,$y$-plane (Figure \ref{fig:flow100} and \ref{fig:flow110}).

\subsection{Comparison with Molecular Dynamics Simulation Data}
\label{ssec:compmd}

Next, we return to the averaged pair diffusion expressed by Eq. \ref{eq:fsesh}, whose predictions are shown in Figure \ref{fig:rhor_cb_30ps} and \ref{fig:rhor_et_100ps} as gray dashed curves. 
Note that at this stage the viscosity entering the prefactor of Eq. \ref{eq:fsesh} has been treated as an empirical fit parameter (see section \ref{ssec:decay} for a detailed discussion of the role of $\eta$). 
Moreover, a single fit curve only has been determined based on the average of all three MD data curves for a given electrolyte in Figure \ref{fig:rhor} (i.e. $\rho_{++}$, $\rho_{+-}$ and $\rho_{--}$). 
We note an excellent agreement for all pair types for intermediate and large separations ($r\gtrsim 10$ \AA), underscoring the universal behavior of the pair diffusion for these distances. 
Obviously, deviations at short $r$ due to the finite size of the ions and their chemical structure, as well as the oscillating deviations for the IL at somewhat larger distances are not captured by the analytical prediction due to the assumption of point particles in our theory. 
Interestingly, similar deviations from ideal behavior have been found for simple hard-sphere fluids \cite{mittal2012pair}, which have been rationalized by an effective diffusive motion of the particles on a free energy landscape imposed by the local structure of the liquid \cite{hummer2005position}. 
By incorporating the Rotner-Prager tensor \cite{rotne1969variational,beenakker1986ewald}, finite ion radii could be captured, although accounting for the local ordering appears to be more challenging. 

To further probe the impact of the local ordering on the distance dependence observed in Figure \ref{fig:rhor}, we computed the differences between the $\rho_{XY}$ determined from the MD data and the analytical prediction of Eq. \ref{eq:fsesh}, i.e. $\Delta\rho_{XY} = \rho_{XY} - \rho_\mathrm{PBC}$, and compared the resulting differences to the appropriately scaled and shifted radial distribution functions (RDFs, dashed curves). 
From Figures \ref{fig:drhor_cb_30ps} and \ref{fig:drhor_et_100ps}, we observe a good agreement of the peak positions of the RDFs and $\Delta\rho_{XY}$ for intermediate $r$, indicating that the deviations of the MD data from Eq. \ref{eq:fsesh} largely arise from the local electrolyte structure. 
Notably, for both the CE and the IL a qualitative agreement between the RDF peaks and $\Delta\rho_{XY}$ is even found for the first solvation shell ($r\approx 5$ \AA), although the deviations are somewhat larger for the CE due to its sharp first cation-anion coordination peak in the RDF. 
Apparently, the ions in the IL can be reasonably approximated as spherical particles, while this  simplification breaks down for the CE because of the preferential coordination of the small lithium ions to the TFSI oxygen atoms. 

Nevertheless, the overall agreement between the MD data and Eq. \ref{eq:fsesh} is fairly good, demonstrating that hydrodynamic interactions significantly govern the cooperative charge transport at larger $r$. 
Importantly, these findings therefore demonstrate that 
not only momentum conservation may lead to anticorrelated motion of ions in periodic systems, but also the approximate incompressibility of the electrolyte giving rise to hydrodynamic interactions. 
Of course, momentum conservation is present in real systems (hence the exclusion of the zeroth mode in Eq. \ref{eq:tfourier}), and clearly affects charge transport in ILs, which we will discuss in section \ref{ssec:conductivity}. 
Strikingly, qualitatively similar features as in Figure \ref{fig:rhor} have even been observed for polymer electrolytes \cite{muller1994permeation,muller1995computer,maitra2008understanding} (although not the main focus of these studies), again underscoring the universality of hydrodynamic interactions, in line with other analytical calculations \cite{farago2011anomalous}.

\section{Time Dependence of Ionic Pair Diffusion}
\label{sec:timedep}

\subsection{Decay of Hydrodynamic Interactions}
\label{ssec:decay}

So far, diffusive dynamics has been implicitly assumed via Eq. \ref{eq:dcross}. 
However, the dynamics is still subdiffusive on a time scale of $\Delta t = 30$ ps (CE) and $100$ ps (IL), for which the $\rho_{XY}$ in Figure \ref{fig:rhor} have been computed. 
Nonetheless, the agreement between the simulation data and Eq. \ref{eq:fsesh} is already almost quantitative. 
Of course, the transport properties introduced in section \ref{sec:intro} are usually evaluated for sufficiently large $\Delta t$, i.e., the diffusive regime. 
However, as argued in section \ref{ssec:theory}, the $\rho_{XY}$-curves become blurred for large $\Delta t$ due to the fact that the distances between the ions change as $\Delta t$ increases (in fact, for $\Delta t\rightarrow\infty$, $\rho_{XY}$ would even converge to a constant value irrespective of $r$). 
On the other hand, for short $\Delta t$, $\rho_{XY}$ is well-defined but the dynamics is still subdiffusive, preventing the evaluation of the contribution of $\rho_{XY}$ to $\sigma_{XY}$. 
Therefore, we characterize the time dependence of $\rho_{XY}$ in more detail in a next step. 

\begin{figure*}
 \subfloat[\label{fig:rhor-vs-t_cb}]{
 \includegraphics[width=0.5\textwidth]{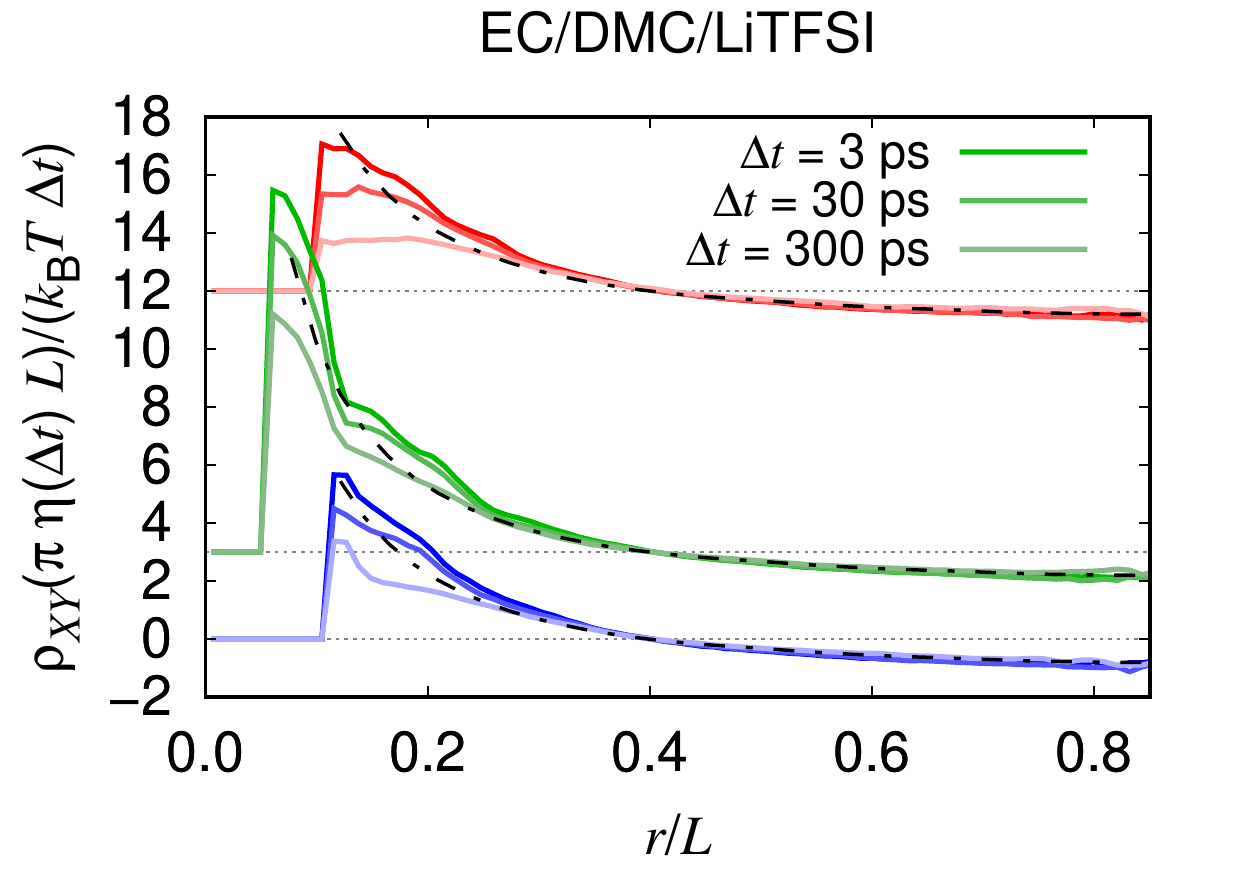}}
 \subfloat[\label{fig:rhor-vs-t_et}]{
 \includegraphics[width=0.5\textwidth]{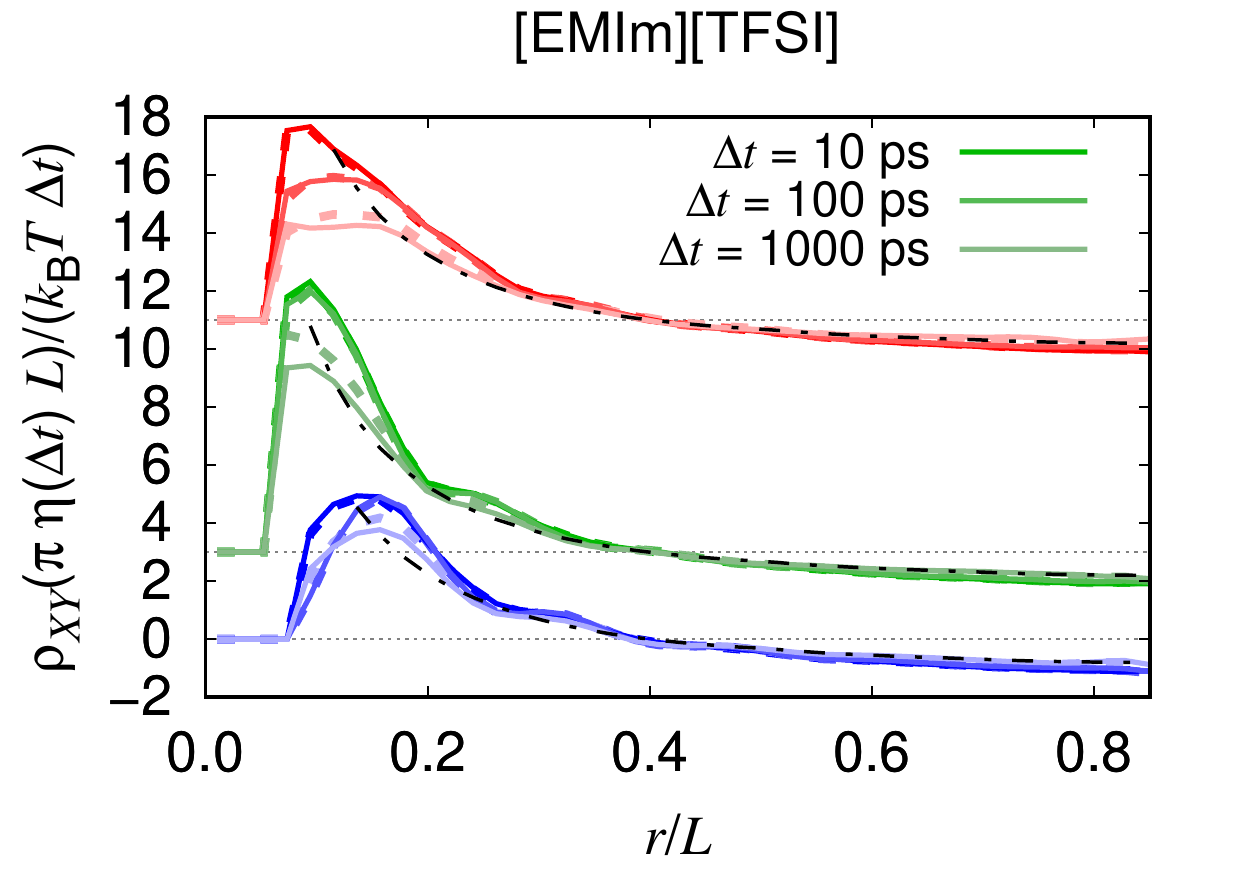}}
\\
 \subfloat[\label{fig:4pir2rhor-vs-t_cb}]{
 \includegraphics[width=0.5\textwidth]{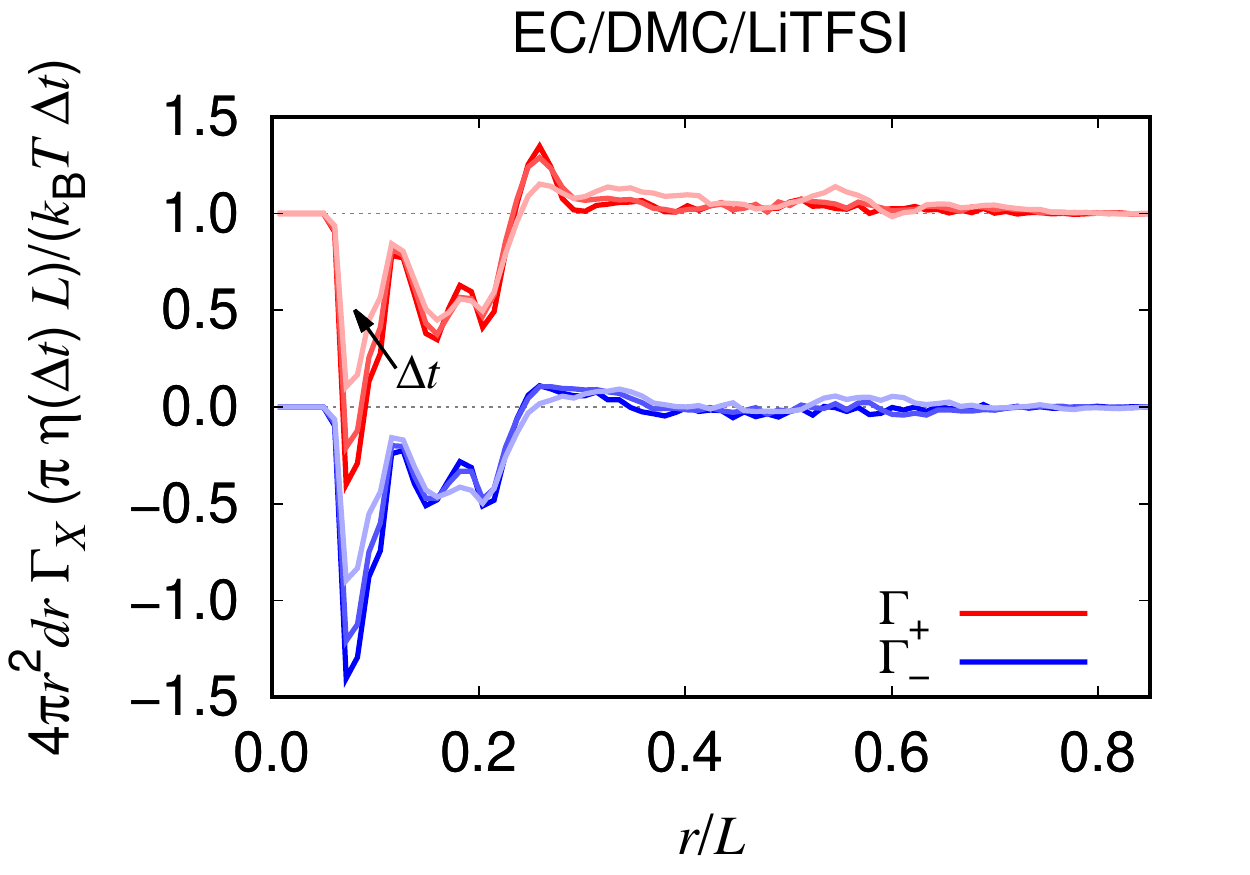}}
 \subfloat[\label{fig:4pir2rhor-vs-t_et}]{
 \includegraphics[width=0.5\textwidth]{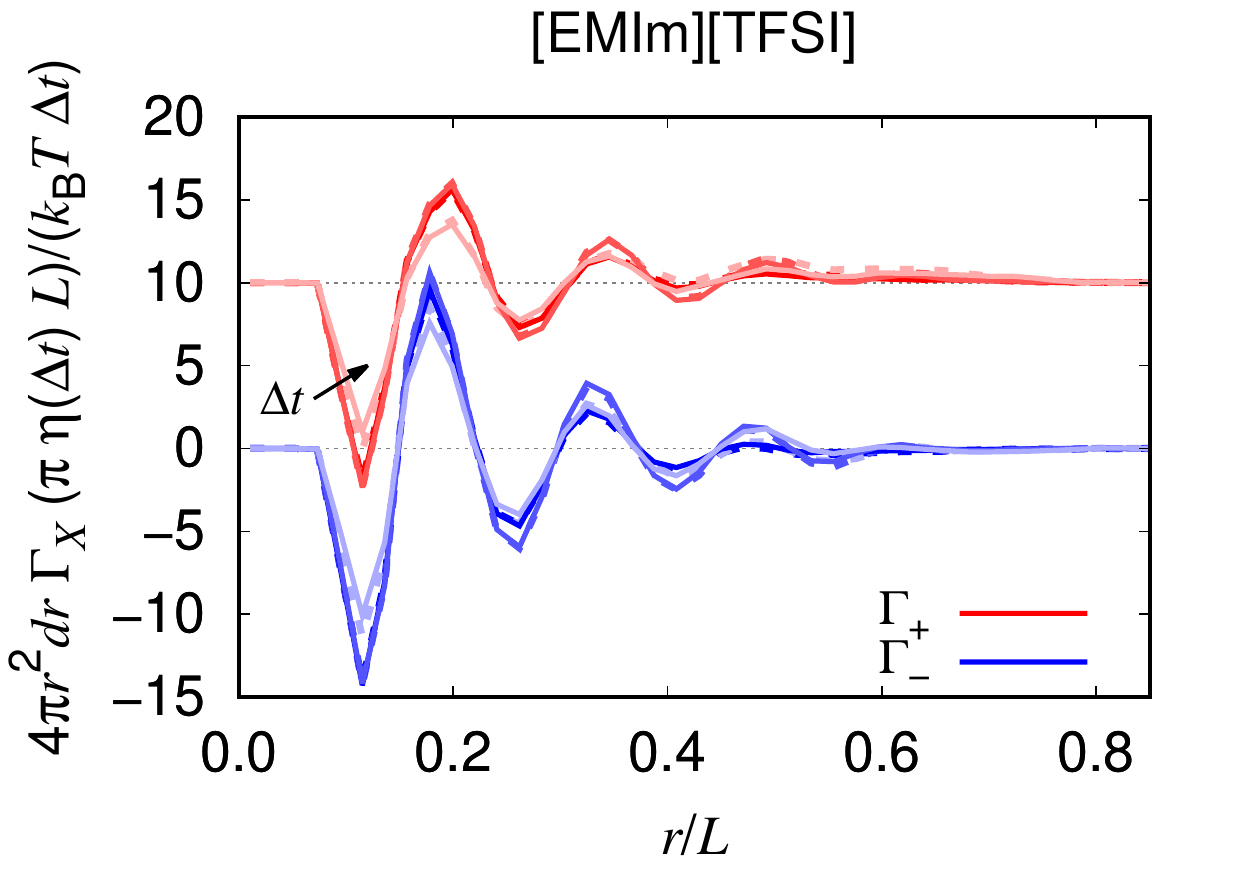}}
\\
 \subfloat[\label{fig:chir-vs-t_cb}]{
 \includegraphics[width=0.5\textwidth]{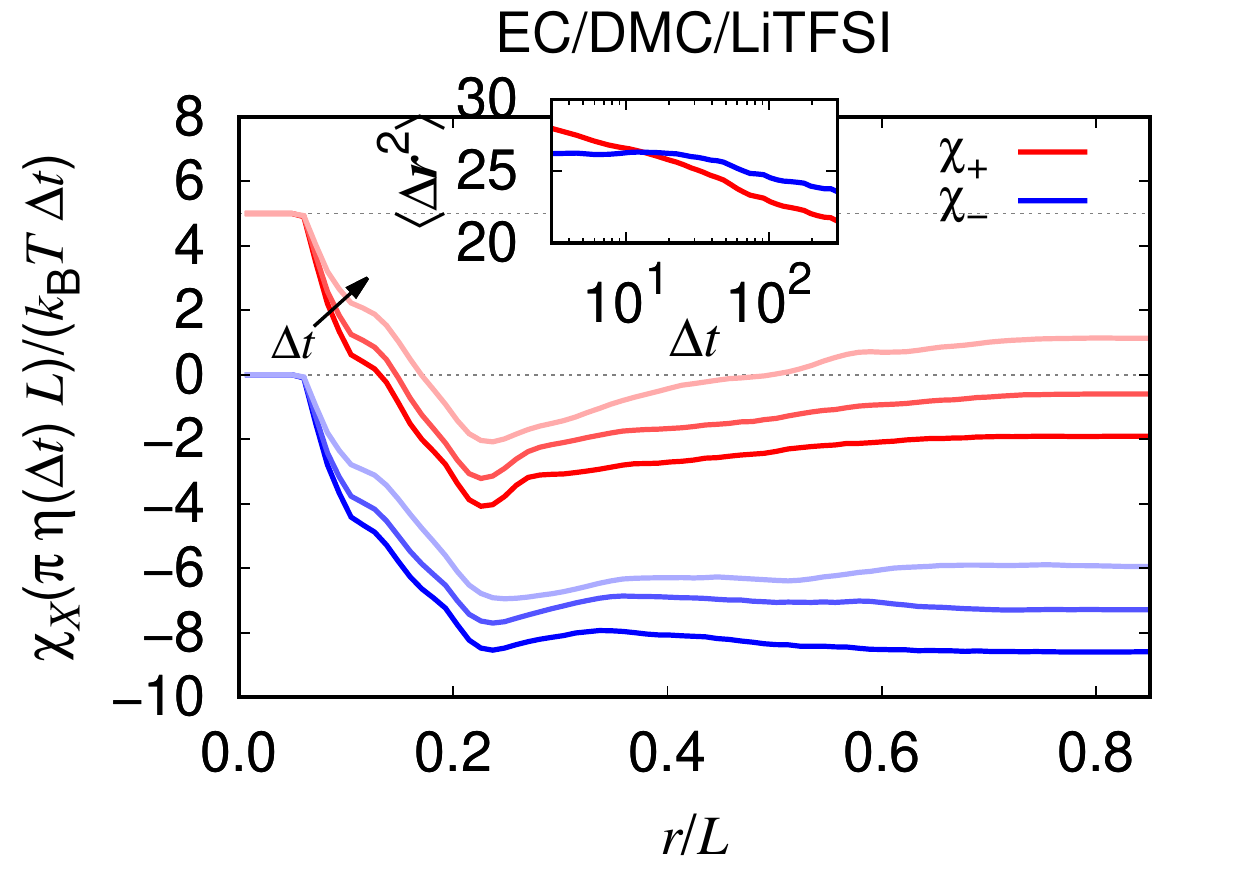}} 
 \subfloat[\label{fig:chir-vs-t_et}]{
 \includegraphics[width=0.5\textwidth]{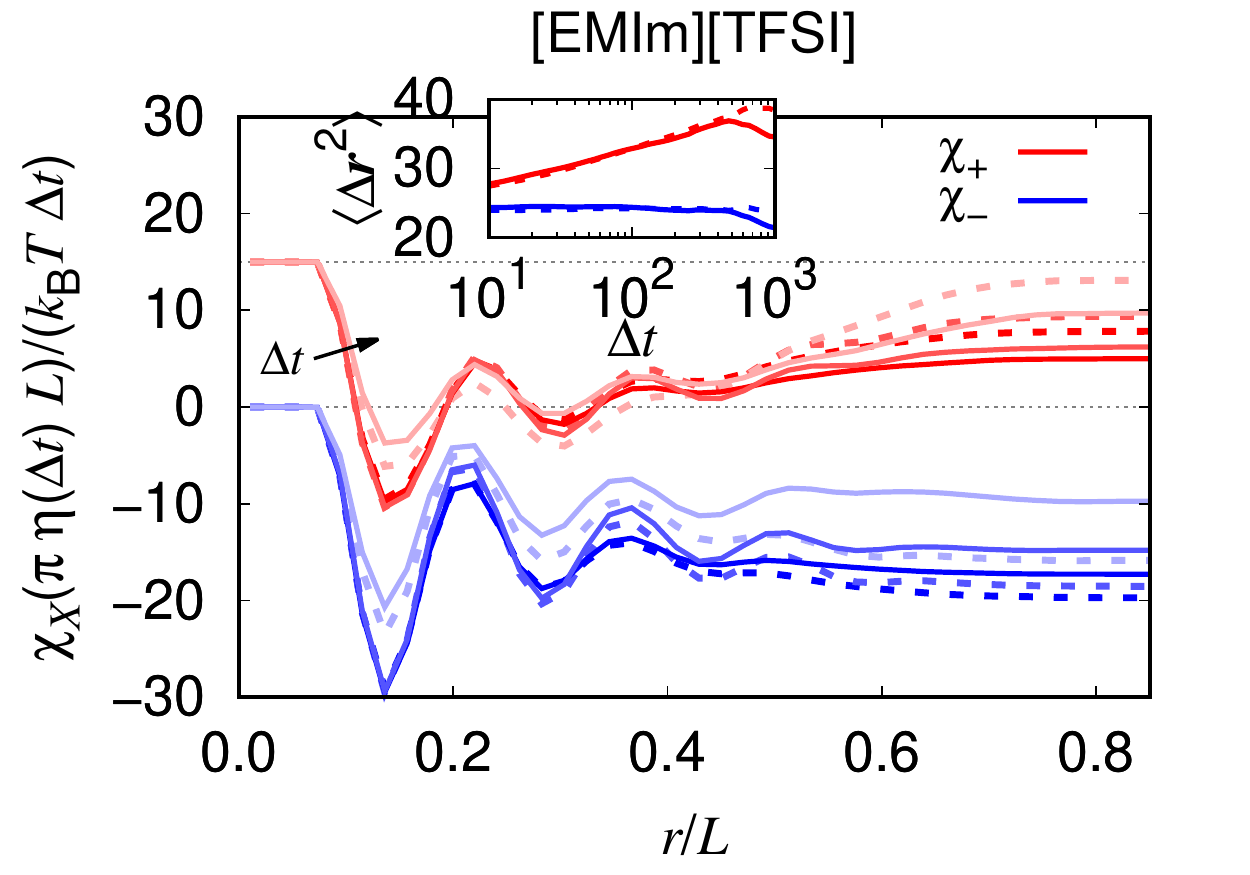}} 
\caption{\label{fig:rhor-vs-t} (a), (b): Normalized ion correlations $\rho_{XY}$ at different $\Delta t$ as a function of normalized distance $r/L$. 
The color coding of the pair types is identical to Figure \ref{fig:rhor}. 
The dash-dotted black curves correspond to the prediction from Eq. \ref{eq:fsesh} with time-dependent viscosity $\eta(\Delta t)$ (see text). 
(c), (d): Weighted and normalized cross-correlations in the integrand $4\pi r^2 \Gamma_\pm$ of Eq. \ref{eq:chi} (see text). 
(e), (f): Integrated and normalized cross-correlations $\chi_+ = \chi_{++} - \chi_{+-}$ and $\chi_- = \chi_{+-} - \chi_{--}$. 
The MSDs in the same normalization are shown in the insets of (e) and (f). 
The dashed lines in (b), (d) and (f) correspond to the ILs with modified masses. 
The relative uncertainties of the unnormalized plateau values at the largest $\Delta t$ in (e) and (f) are in the range of $10$ \%. 
Several curves have been shifted for better visibility.}
\end{figure*}

Naturally, for larger $\Delta t$, the magnitude of $\rho_{XY}$ is larger in the simulations because the ions traveled larger distances on average. 
Theoretically, space and time dependence of $\rho_{XY}$ are separated in Eq. \ref{eq:fsesh} as the former is given by the universal function $[(L/r)-\xi(L/r)]$, whereas the time dependence is exclusively contained in the prefactor. 
Although not immediately apparent, not only the linear $\Delta t$-term in the prefactor of Eq. \ref{eq:fsesh}, but also the short-time viscosity $\eta(\Delta t)$, which is not fully converged in the subdiffusive regime, lends $\rho_{XY}$ its time dependence. 
In analogy to Eq. \ref{eq:gketa}, we therefore phenomenologically define the short-time viscosity as 
\begin{equation}
 \label{eq:gketat}
 \eta(\Delta t) = \frac{V}{k_\mathrm{B}T} \int_0^{\Delta t} dt \langle P_{\beta\gamma}(0)P_{\beta\gamma}(t)\rangle\text{.}
\end{equation}
Details of the calculation of the time-dependent viscosity are given in Appendix \ref{sec:visc}. 
Figures \ref{fig:rhor-vs-t_cb} and \ref{fig:rhor-vs-t_et} show $\rho_{XY}$ for for three different $\Delta t$-values normalized by $(k_\mathrm{B}T\Delta t)/(\pi \eta(\Delta t) L)$ as expected from Eq. \ref{eq:fsesh} to yield dimensionless quantities. 
In this way, $\rho_{XY}$ becomes susceptible to dynamical intricacies beyond its trivial dependence on the viscosity. 
In other words, one can assess whether space and time dependence are strictly separated suggested by Eq. \ref{eq:fsesh}. 
Note that the largest $\Delta t$-values in Figure \ref{fig:rhor-vs-t} were chosen such that they are slighly smaller than the onset of the diffusive regime (see below) for reasons mentioned above. 
In particular, we choose $\Delta t = 3$, $30$ and $300$ ps for the CE and $\Delta t = 10$, $100$ and $1000$ ps for the IL. 

From Figures \ref{fig:rhor-vs-t_cb} and \ref{fig:rhor-vs-t_et} we observe the characteristic hydrodynamic behavior for all $\Delta t$, that is, both on time scales close to the diffusive regime and within the subdiffusive regime down to a few picoseconds. 
However, with increasing $\Delta t$, both the locally positive ($\rho_{XY}>0$) and globally negative correlation ($\rho_{XY}<0$) dimishes in the normalized representation. 
This can be attributed to the fact that after a certain time, the initial interionic distances $r$ change due to the motion of the ions, resulting in exchange processes within their coordination shells. 
Consequently, the initial $\rho_{XY}$-values become averaged over different $r$, leading to the observed decay. 
This is in agreement with the average distances traveled by the ions during the individual $\Delta t$-values as estimated from the mean squared displacements (MSDs). 
In particular, for the CE we find displacements (averaged over both ion species) of $1.2$, $2.5$ and $6.6$ \AA\ for the respective $\Delta t$, whereas for the IL the corresponding values are $1.2$, $2.2$ and $5.3$ \AA, the values for the largest $\Delta t$ being comparable to the ion sizes. 
As mentioned above, for $\Delta t \rightarrow \infty$, $\rho_{XY}$ would fully decay due to the complete loss of information on the original distances between the different ions. 
Note that for ILs, this constant would be negative due to momentum conservation (section \ref{sec:intro}). 
Interestingly, in the opposite limit $\Delta t \rightarrow 0$, the velocities should obey the Maxwell-Boltzmann distribution, such that the velocities/displacements of two ions are uncorrelated irrespective of their distance. 
Thus, also in the short-time limit, $\rho_{XY}$ would be a constant (which is negative for ILs due to residual correlations satisfying momentum conservation \cite{tu2014spatial2}). 
However, from Figures \ref{fig:rhor-vs-t_cb} and \ref{fig:rhor-vs-t_et} we find that starting from a few picoseconds, the hydrodynamic picture already holds. 
This is also reflected by the comparison with the theoretical curve (Eq. \ref{eq:fsesh}, black dash-dotted lines in Figures \ref{fig:rhor-vs-t_cb} and \ref{fig:rhor-vs-t_et}; unlike in Figure \ref{fig:rhor}, the curves were only normalized, but no fitting was performed). 
For both electrolytes, we find that for short $\Delta t$ the magnitude of $\rho_{XY}$ is larger than the prediction by about $10-30$ \% for all except short $r$, for which the comparison breaks down due to the local structure. 
Keeping in mind that the dynamics only becomes diffusive at the respective largest $\Delta t$, larger $\rho_{XY}$-values in the MD simulations are not surprising as it is generally observed that when applying Eq. \ref{eq:dself} (or the corresponding equation for pair diffusion) in the subdiffusive regime, the resulting approximate (pair) diffusion coefficients are larger than their long-time values (see Figure \ref{fig:sig-vs-t} in section \ref{ssec:selfnx} below). 
However, as argued above, $\rho_{XY}$ decays due to structural rearrangements in the opposite limit $\Delta t\rightarrow\infty$. 
In this context, it is important to stress a subtle difference between the definition of $\rho_{XY}$ as extracted from the MD data and the hydrodynamic theory in section \ref{ssec:theory2}: 
Eq. \ref{eq:fsesh} was derived on the basis of a continuous flow field (see for example Eq. \ref{eq:dgk} or Eq. \ref{eq:stokes}), therefore, the pair diffusion coefficient is determined for a \emph{fixed} distance $r$ between two points within the periodic cell. 
On the contrary, the interionic distance in the MD simulations evolves with time, which ultimately leads to the complete decay of $\rho_{XY}$. 
Implications of this conceptual difference can indeed be observed from Figures \ref{fig:rhor-vs-t_cb} and \ref{fig:rhor-vs-t_et}: 
While Eq. \ref{eq:fsesh} predicts that space and time dependence can be separated, the MD curves still show a residual time dependence (i.e. their decay) despite the normalization by the prefactor of Eq. \ref{eq:fsesh}, which theoretically contains the entire time dependence. 
At the onset of the diffusive regime, the average displacement of the ions becomes comparable to their own size or the size of their solvation shell \cite{self2019transport,wettstein2022controlling} ($\sim 5$ \AA), which is still small compared with the large and intermediate distances in Figures \ref{fig:rhor-vs-t_cb} and \ref{fig:rhor-vs-t_et}. 
Consequently, we observe that the hydrodynamic picture is still valid for the largest $\Delta t$-values before the trivial long-time decay continues. 
For the largest $\Delta t$-values, we find that the MD curves in Figures \ref{fig:rhor-vs-t_cb} and \ref{fig:rhor-vs-t_et} match the prediction from Eq. \ref{eq:fsesh} almost quantitatively for larger $r$. 
Several factors could contribute to this conincidence: 
First, the pair diffusion coefficients are no longer overestimated as in the subdiffusive regime, second, Eq. \ref{eq:gketat} converges to the constant long-time viscosity and thus becomes equivalent to Eq. \ref{eq:gketa}, and finally, the ion displacements are still comparatively small to the interionic distances on which hydrodynamic interactions are relevant. 

It is also worth noting that the $\rho_{XY}$ are similar to the coupling factor $\lambda$ defined in our recent work on IL/Li-salt mixtures \cite{wettstein2022controlling}. 
In particular, $\lambda$ expresses the degree of coupled diffusion between the displacement vectors of initially neighbored ions, similar in spirit to a correlation coefficient. 
For IL/Li-salt mixtures, we observed a large degree of coupled diffusion for small salt concentrations ($\lambda\approx 0.8$) due to the stable lithium coordination shell composed of anions \cite{wettstein2022controlling}. 
Conversely, substantially smaller values for $\lambda$ were found at high concentrations because anions are shared between distinct lithium ions as coordination partners. 
Consistent with the alternating structure of cations and anions in pure ILs studied in this work, leading to shared coordination partners as well, one would expect a moderate albeit significant degree of coupled diffusion. 
Indeed, when normalizing $\rho_{XY}$ by the MSDs in analogy to $\lambda$, we find values $0.35$ for the first coordination sphere of the IL. 
For the CE, this value is significantly larger ($\sim 0.6$) as expected from the lower ion concentration such that shared coordination shells composed of one anion and two cations (or vice versa) hardly emerge. 
This is also in line with a recent analysis showing that in CEs the transport mainly occurs in a vehicular fashion, i.e. collectively with the local enviroment \cite{andersson2022dynamic}. 
Moreover, in our previous work we found that $\lambda$ decays with increasing time due to the fact initially nearby ions disengange \cite{wettstein2022controlling}, compatible with the present observations from Figure \ref{fig:rhor-vs-t_cb} and \ref{fig:rhor-vs-t_et}. 
Nonetheless, we found previously that even after a neighboring ion left a given ion's coordination sphere, some residual dynamical coupling persists as a result of the hydrodynamic flow field \cite{wettstein2022controlling}. 
The hydrodynamic theory developed in the preceding section fully rationalizes these earlier findings, as it accounts for any interionic distance. 

Finally, Figure \ref{fig:rhor-vs-t_et} also shows the respective normalized $\rho_{XY}$-curves for the systems in which the ion masses have been scaled (section \ref{sec:sim}) as dashed lines. 
While the differences as compared to the original systems appear to be minute for $\rho_{XY}$, they will turn out to be crucial for the contributions to $\sigma_{XY}$ (section \ref{ssec:conductivity}).

\subsection{Implications for the Conductivity}
\label{ssec:conductivity}

Next, we discuss how the electrolyte structure, the hydrodynamic flow field as well as the deviations from it govern $\sigma_+$ and $\sigma_-$. 
More generally, to relate $\rho_{XY}$ and $\sigma_{XY}$, we express Eq. \ref{eq:sigij} as 
\begin{widetext}
\begin{equation}
 \label{eq:sum2int2}
 \sigma_{XY}(\Delta t) = \frac{e^2 N_X}{6 V\, k_\mathrm{B}T \Delta t}\bigg[\langle\Delta{\bf r}^2(\Delta t)\rangle_X\,\delta_{XY} + \frac{z_X z_Y(N_Y-\delta_{XY})}{V}\int_V\,d{\bf r}\,\rho_{XY}(r,\Delta t)\,g_{XY}(r)\bigg]
\end{equation}
\end{widetext}
in analogy to early analytical work \cite{fuoss1963conductance,lee1978conductance,ebeling1979electrolytic,altenberger1983theory,yamaguchi2009theoretical} and recent simulation studies \cite{tu2014spatial,tu2014spatial2,matubayasi2019spatial}. 
Here, $g_{XY}(r)$ denotes the RDF between ion species $X$ and $Y$, $\langle\Delta{\bf r}^2(\Delta t)\rangle_X$ is the MSD of species $X$ during $\Delta t$ and $\delta_{XY}$ is the Kronecker delta. 
For $\sigma_X = \sigma_{XX} + \sigma_{XY}$, we rewrite Eq. \ref{eq:sigpm} as 
\begin{equation}
 \label{eq:sum2int3}
 \sigma_X(r,\Delta t) = \frac{e^2 N_X}{6 V k_\mathrm{B}T \Delta t}\left[\langle\Delta{\bf r}^2(\Delta t)\rangle_X + \chi_X(r,\Delta t)\right]\text{,}
\end{equation}
where the first term on the right-hand side is the ideal Nernst-Einstein conductivity 
\begin{equation}
 \label{eq:sig0}
 \sigma_{X,0}(\Delta t) = \frac{e^2 N_X D_X(\Delta t)}{V\,k_\mathrm{B}T}\text{,}
\end{equation}
arising from the self-diffusion of species $X$ (Eq. \ref{eq:dself}) and 
\begin{widetext}
\begin{equation}
 \label{eq:chi}
 \chi_X(r,\Delta t) = \int_0^r\,dr'\,4\pi r'^2\,\frac{N_X}{V}\left[\frac{N_X-1}{N_X}\rho_{XX}(r',\Delta t)\,g_{XX}(r')-\rho_{XY}(r',\Delta t)\,g_{XY}(r')\right] = \int_0^r\,dr'\,4\pi r'^2\,\Gamma_X(r',\Delta t)
\end{equation}
\end{widetext}
is a short-hand notation for the distance-dependent cross correlations experienced by ions of the type $X$. 
In Eq. \ref{eq:chi}, we expressed $\chi_X$ as a function of the upper bound $r$ of the integral. 
Although spherical integration can be carried out due to isotropy, it should be emphasized that the integral in Eq. \ref{eq:sum2int2} also contains contributions for $L/2<r\leq \sqrt{3}L/2$, for which $g_{XY}(r) < 1$. 

For an ideal structureless electrolyte with $g_{XY}(r)=1$ and $\rho_{XY}$ strictly given by Eq. \ref{eq:fsesh} (i.e. no deviations as observed in Figure \ref{fig:rhor}), $\chi_X$ would be zero when integrated over the entire box because the integral over Eq. \ref{eq:dconv} vanishes \cite{figueirido1995finite,hummer1998molecular}. 
However, even for a real electrolyte it is obvious from Figures \ref{fig:rhor-vs-t_cb}, \ref{fig:rhor-vs-t_et} and Eq. \ref{eq:chi} that when calculating $\chi_+$ and $\chi_-$, the hydrodynamic interactions contained in, say, $\rho_{++}$ will largely cancel with those of $\rho_{+-}$. 
That is, only when either $\rho_{XY}$ is non-ideal (Figure \ref{fig:rhor}) or when $g_{XY}\neq 1$, remaining contributions to $\chi_\pm$ can be expected when subtracting the two integrands in Eq. \ref{eq:chi}. 
Before embarking on the discussion of $\chi_\pm$, it is therefore instructive to consider the difference of the two integrands in Eq. \ref{eq:chi}, denoted as $\Gamma_\pm$ and weighted by $4\pi r^2$ due to radial symmetry (Figures \ref{fig:4pir2rhor-vs-t_cb} and \ref{fig:4pir2rhor-vs-t_et}; the same normalization as for $\rho_{XY}$ has been applied). 
We find that for both the CE and the IL, the non-vanishing non-hydrodynamic contribution at short distances is negative (i.e. decreasing the overall conductivity) due to the preferential interactions of oppositely charged ions on a local scale. 
However, while for the CE only marginal contributions are observed for $r/L\gtrsim 0.3$ (Figure \ref{fig:4pir2rhor-vs-t_cb}), several additional peaks occur up to a distance of $r/L\gtrsim 0.6$ for the IL (Figure \ref{fig:4pir2rhor-vs-t_et}). 
As expected from the decay of $\rho_{XY}$ for larger $\Delta t$, the magnitude of the non-vanishing cross-correlations decreases for both the CE and the IL. 
The dashed curves in Figure \ref{fig:4pir2rhor-vs-t_et} again show the results for the IL with modified ion masses. 
As in Figure \ref{fig:rhor-vs-t_et} these differences appear to be minute, but will turn out to be significant upon integration via Eq. \ref{eq:chi}, which we study next. 

Figures \ref{fig:chir-vs-t_cb} and \ref{fig:chir-vs-t_et} show the integrated $\chi_\pm$ as a function of the upper bound of the integral in Eq. \ref{eq:chi} with the same normalization as before. 
As already expected from Figures \ref{fig:4pir2rhor-vs-t_cb} and \ref{fig:4pir2rhor-vs-t_et}, the negative contribution at short distances is dominating, such that the entire integral is smaller than zero for all $r$ (note that the $\chi_+$-curves have been shifted for clarity). 
While this behavior is encountered for both electrolytes, only the IL shows significant contributions beyond local scales, again reflected by multiple peaks arising from the rather long-ranged ordering. 
In contrast, $\chi_-$ is approximately constant for $r/L \gtrsim 0.4$ in case of the CE, whereas $\chi_+$ still displays minor changes for larger $r/L$, probably due to a minor ordering of the ions on these scales. 
Another interesting IL-specific effect can be observed at large $r$ from Figure \ref{fig:chir-vs-t_et}: 
Here, $\chi_+$ still changes slighly for $r/L > 0.6$. 
By comparison with the curves of the IL with modified masses (dashed curves), it becomes obvious that this is an imprint of the momentum-conservation constraint, as the the correlated ionic motion at more local scales has to be globally compensated (section \ref{sec:intro}). 
A similar observation can be made for the second IL studied here, [EMIm][BF$_4$], which is shown in Appendix \ref{sec:emimbf4} (Figure \ref{fig:chir-vs-t_eb}). 
With increasing $\Delta t$, the magnitude of the peaks of $\chi_\pm$ decreases for both CE and IL due to local relaxation processes as already observed from Figures \ref{fig:4pir2rhor-vs-t_cb} and \ref{fig:4pir2rhor-vs-t_et}. 

Interestingly, also the plateau values of $\chi_\pm$ at $r/L\rightarrow\sqrt{3}/2$, reflecting the overall cross-correlations, decrease with increasing $\Delta t$, demonstrating that not all dynamical features in the subdiffusive regime are captured by $\eta(\Delta t)$. 
A similar finding was already made in context of Figures \ref{fig:rhor-vs-t_cb} and \ref{fig:rhor-vs-t_et}: 
Unlike the hydrodynamic theory based on a continuous flow field, the distances between discrete ions in the simulation relax with time, which additionally contributes to their mutual pair diffusion. 
Apparently, the residual time dependence of $\rho_{XY}(\Delta t)$ not contained in $\eta(\Delta t)$ (Figures \ref{fig:rhor-vs-t_cb} and \ref{fig:rhor-vs-t_et}) does not entirely vanish when performing the integration according to Eq. \ref{eq:chi} despite locally positive and globally negative correlations. 
In particular, for the CE the magnitude of $\chi_+^\mathrm{int} = \chi_+(r/L\rightarrow\sqrt{3}/2)$ decreases by about $40$ \% and the corresponding $\chi_-^\mathrm{int}$ decreases by a comparable amount of $30$ \% when going from $3$ to $300$ ps. 
Due to the fact that $\chi_\pm^\mathrm{int}$ is negative and hence decreasing both $\sigma_\pm$ and $\sigma$, its decrease in magnitude because of the additional relaxation processes enhances the conductivity. 
In other words, if the dynamics was entirely governed by $\eta(\Delta t)$, $\chi_\pm^\mathrm{int}$ would remain constant, leading to lower $\sigma$-values for longer $\Delta t$. 
However, the decrease of $|\chi_\pm^\mathrm{int}|$ is overcompensated by an increase of $\eta(\Delta t)$ by $130$ \% between $3$ and $300$ ps, contributing to the fact that $\sigma(\Delta t)$ decreases with increasing $\Delta t$ in the subdiffusive regime (see below). 
For the IL, the magnitude of $\chi_+^\mathrm{int}$ and $\chi_-^\mathrm{int}$ decreases by $40-50$ \% when going from $10$ to $1000$ ps, whereas $\eta(\Delta t)$ increases by a factor of almost six. 
This indicates that although relaxation of the electrolyte structure affects $\chi_\pm^\mathrm{int}$ and thus the conductivity value, the high viscosity of ILs slows down the ionic motion more strongly as compared to other electrolytes. 

In case of ILs, the $\Delta t$-dependence of $\chi_\pm^\mathrm{int}$ is influenced by momentum conservation. 
This is best seen from the comparison of [EMIm][TFSI] and [EMIm][BF$_4$]: 
For [EMIm][TFSI], the value decreases for $\chi_+^\mathrm{int}$ and $\chi_-^\mathrm{int}$ with increasing $\Delta t$, both with standard and with modified masses (Figure \ref{fig:chir-vs-t_et}). 
While the same behavior is found for [EMIm][BF$_4$] for standard masses, the trend becomes reverted for $\chi_+^\mathrm{int}$ when the masses are scaled (Figure \ref{fig:chir-vs-t_eb}). 
In this context, it is noteworthy that by our scaling procedure, the anions become heavier than the cations for [EMIm][BF$_4$], while the opposite is true for standard molar masses. 
For [EMIm][TFSI], however, the cations are lighter than the anions in both cases. 
One may therefore speculate that the compensation of the local exchange processes by the motion of remote ions is affected by these details.

\subsection{Self-Diffusion and Cross-Correlation}
\label{ssec:selfnx}

\begin{figure*}
 \subfloat[\label{fig:sig-vs-t_cb}]{
 \includegraphics[width=0.5\textwidth]{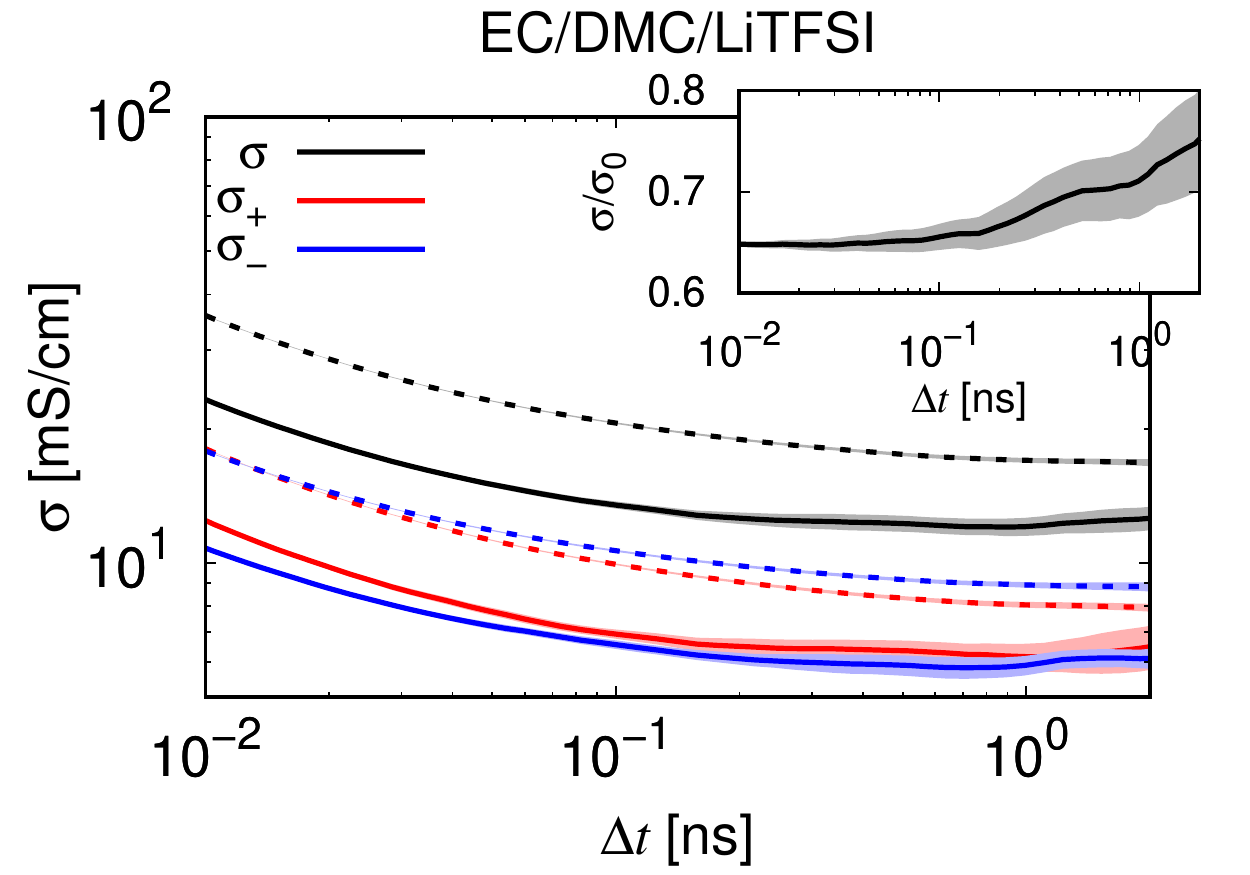}} 
 \subfloat[\label{fig:sig-vs-t_et}]{
 \includegraphics[width=0.5\textwidth]{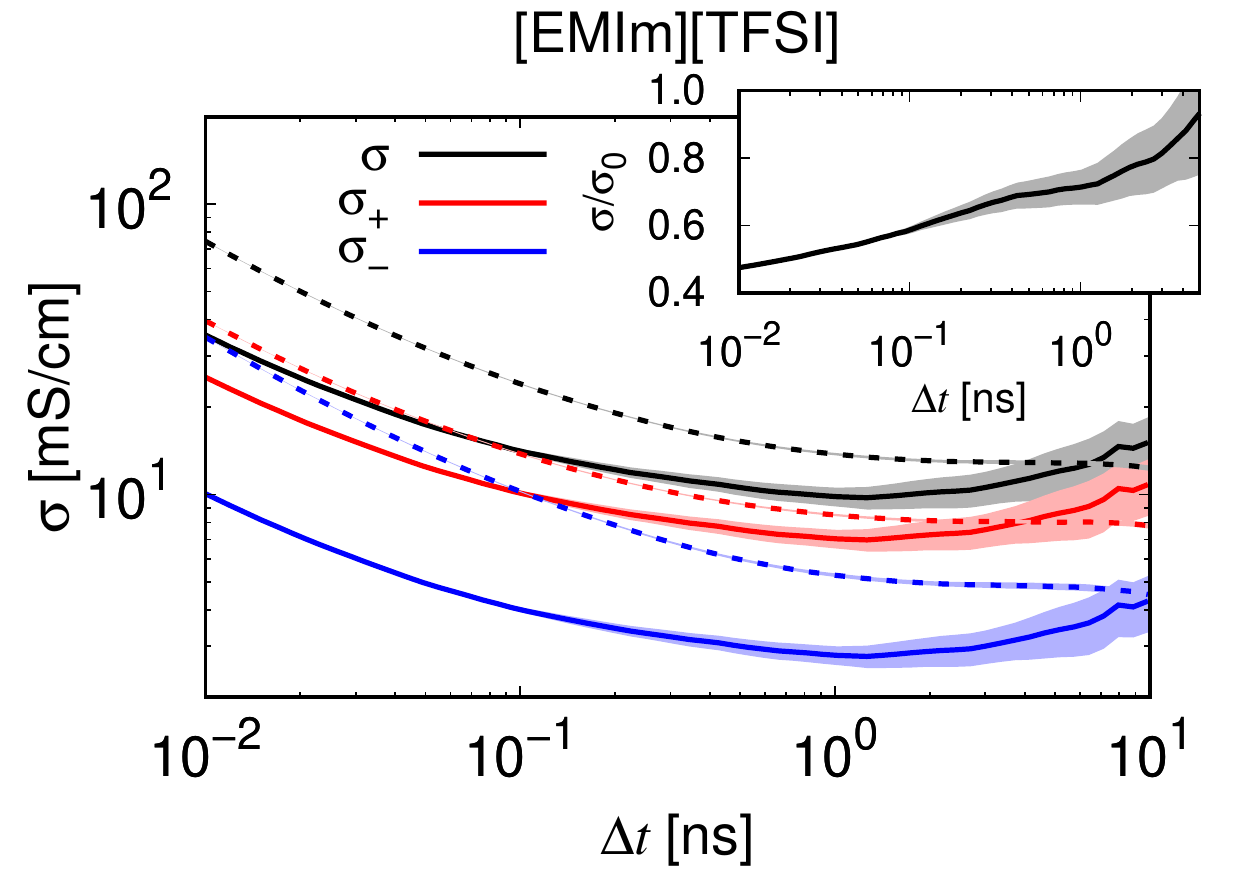}}
\caption{\label{fig:sig-vs-t} Time dependent conductivities (solid lines) and ideal Nernst-Einstein conductivities (dashed lines) for (a) the CE and (b) the IL. The standard deviations are indicated as shaded regions. The degree of uncorrelated ion motion, expressed as $\sigma/\sigma_0$, is shown in the insets.}
\end{figure*}

Naturally, apart from the integrated cross-correlations $\chi_\pm^\mathrm{int}$, the self-diffusion of the ions contributes a large fraction to the total conductivity. 
In practice, one therefore usually aims to either increase the mobility of the ions (often the cation), but also to alter the ionic correlations by employing different salts or solvents in order to optimize an electrolyte \cite{xu2004nonaqueous,xu2014electrolytes}. 
The insets in Figures \ref{fig:chir-vs-t_cb} and \ref{fig:chir-vs-t_et} show the ions' MSDs with the same normalization as for $\chi_\pm^\mathrm{int}$, making both quantities directly comparable. 
As for the cross-correlations, we observe that not all dynamical features affecting the subdiffusive regime are captured by $\eta(\Delta t)$, reflected by an additional $\Delta t$-dependence in Figures \ref{fig:chir-vs-t_cb} and \ref{fig:chir-vs-t_cb}. 
In particular, both curves become slightly smaller for the CE, whereas in case of the IL the cationic contribution increases while the anionic contribution remains constant. 
This shows that at least for certain ionic species, additional short-time processes, e.g. arising from local relaxation processes, the ions' internal degrees of freedom or forward-backward correlations, affect the MSDs in a different fashion than the collective property $\eta(\Delta t)$. 
However, this apparent deviation from the simplified Stokes-Einstein relation is not too surprising. 
From experimental work and simulations it is known that the Stokes-Einstein relation provides a reasonable first estimate, although deviations of about a factor of two are commonly observed \cite{kaintz2013solute,andersson2022dynamic}. 
Interestingly, however, both the MSD and $\eta(\Delta t)$ remain unaffected within the uncertainties when changing the masses (dashed curves Figures \ref{fig:chir-vs-t_cb} and \ref{fig:chir-vs-t_et} as well as Appendix \ref{sec:visc}), consistent with the overdamped dynamics in a highly-viscous medium that is commonly assumed. 

Finally, Figure \ref{fig:sig-vs-t} shows the conductivities $\sigma_{X}$ together with ideal Nernst-Einstein conductivities $\sigma_{X,0}$ (Eq. \ref{eq:sig0}). 
Importantly, due to momentum conservation, $\sigma_+/\sigma_- = m_-/m_+$ (section \ref{sec:intro}). 
However, as evident from Figures \ref{fig:chir-vs-t_cb} and \ref{fig:chir-vs-t_et}, the product $\chi_\pm^\mathrm{int}(\Delta t)\,\eta(\Delta t)$ is not strictly constant, which implies that also the diffusion coefficients contained in $\sigma_{X,0}$ cannot be fully governed by $\eta(\Delta t)$, as otherwise $\sigma_+/\sigma_- \neq \text{const.}$ (cf. Eq. \ref{eq:sum2int3}). 
Notably, the additional dynamical contributions not contained in $\eta(\Delta t)$ and affecting self-diffusion and cross-correlations in a different way have important technical implications for MD simulations: 
The ratio $\sigma/\sigma_0$, also termed degree of uncorrelated motion, which is frequently used to quantify the deviations from the Nernst-Einstein behavior \cite{borodin2006litfsi,wheatle2018effect,oldiges2018understanding}, itself is time-dependent, at least for the CE and the ILs studied in this work (see insets of Figure \ref{fig:sig-vs-t_cb} and \ref{fig:sig-vs-t_et}). 
Due to the fact that the uncertainties of $\sigma_0$ are considerably smaller than those of $\sigma$ (the former can be averaged over all $N$ ions in the system, leading to uncertainties roughly lower by a factor of $\sqrt{N}$) \cite{muller1995computer,france2019correlations}, one might be tempted to determine $\sigma/\sigma_0$ on short subdiffusive scales and then using this ratio in combination with the statistically more robust $\sigma_0$ to extrapolate $\sigma$ to the diffusive regime. 
However, our observations from Figure \ref{fig:sig-vs-t} show that such an approach is generally not valid because $\sigma/\sigma_0$ varies between $0.5$ and $0.7$ at $\Delta t = 10$ and $1000$ ps, respectively, for the IL. 
For the CE, the variation in $\sigma/\sigma_0$ is smaller albeit significant. 
Therefore, extrapolation from short $\Delta t$ would underestimate the true $\sigma$. 
Consequently, the explicit calculation of $\sigma$ is necessary. 

In total, our findings demonstrate that the pair-diffusion contribution to the conductivity is affected by at least two dynamical features: 
First, hydrodynamic interactions largely govern the overall dynamics of the system. 
Apart from the electrolyte structure affecting the precise value of the hydrodynamic integral, the viscosity is a key parameter characterizing these interactions, in line with the well-known Walden picture \cite{walden1906organische,yoshizawa2003ionic,ueno2010ionicity,lesch2014combined,oldiges2018understanding} and recent MD results \cite{shao2020role}. 
Second, however, relaxation processes, leading to changes in the interionic distances, give rise to additional dynamical contributions which are not captured by the hydrodynamic theory. 
For the electrolytes studied in this work, these deviations lead to an enhancement of the overall conductivity, although it is unclear whether this is generally the case. 
While the importance of hydrodynamic interactions was already recognized in early analytical treatments of ionic conductivity \cite{fuoss1963conductance,lee1978conductance,ebeling1979electrolytic,altenberger1983theory}, the deviations observed in this work are more intricate but can be probed by simulations. 
In this context, it is also noteworthy that it has recently been speculated for CEs that the local viscosity of the enviroment around an ion or a solvate structure rather than the global viscosity is important for diffusion \cite{andersson2022dynamic}. 
Similar local friction effects have been discussed in context of the structural relaxation of ILs \cite{yamaguchi2018coupling,amith2021relationship}. 
It seems plausible that such local viscous effects are relevant for the pair diffusion as well. 
This is even more reasonable as our current theory is based on a single-component fluid with point particles. 
Theories describing hydrodynamic flow in multicomponent systems \cite{wacholder1972slow,wolynes1976slip} or finite ion radii \cite{rotne1969variational,beenakker1986ewald} thus are possible extensions of the model. 
Nevertheless, the present work shows that the hydrodynamic picture holds to a very good approximation until the onset of the diffusive regime. 
In pure-salt electrolytes such as ILs, the pair diffusion is also affected by momentum conservation. 
A similar effect is expected for highly concentrated ternary electrolytes \cite{yamada2019advances,borodin2020uncharted}. 
Through a detailed analysis, the distinct contributions can be disentangled to deliberately optimize electrolytes.

\section{Conclusions and Outlook}
\label{sec:conc}

In this paper, we presented an analytical theory describing the distance dependence of the pair diffusion in periodic systems. 
Essentially, due to the incompressibility of the medium, our theory predicts locally correlated motion, which is compensated by a counterflux at large distances. 
We find a very good agreement between the analytical prediction and the distance dependence of dynamical ion correlations in ILs extracted from MD simulation data, although noticable deviations occur due to several different reasons: 
First, the local structure of the electrolyte and the resulting effective potentials acting on the ions give rise to deviations from the theoretical prediction at short distances. 
Second, on larger time scales, the relaxation of the electrolyte structure leads to the decay of the hydrodynamic interactions. 
Finally, for ionic liquids, the physical constraint of momentum conservation acts on larger length scales. 
Despite this important constraint, our theory shows that the anticorrelated motion occuring for ionic liquids at large distances in periodic systems can be largely rationalized by hydrodynamic interactions arising from the incompressibility of the electrolyte. 
Consequently, the same qualitative behavior is observed for ternary electrolytes. 

The decay of the hydrodynamic interactions is largely governed by the viscosity, in line with the well-known Walden picture \cite{walden1906organische,yoshizawa2003ionic,ueno2010ionicity,lesch2014combined,oldiges2018understanding}. 
However, the relaxation of the electrolyte structure is not captured by the hydrodynamic theory, such that significant deviations arise that affect the ionic cross correlations. 
Nonetheless, the hydrodynamic picture remains valid until the dynamics becomes diffusive. 
Via our framework, it is possible to separate the relative importance of hydrodynamic effects and relaxation, which -- in addition to the electrolyte structure -- govern the collective dynamics between distinct ions. 
Because quantitatively different deviations occur for the self-diffusion, also the degree of uncorrelated motion becomes time-dependent in the subdiffusive regime. 

From the perspective of battery science and electrochemistry, incorporating electrode interfaces into the formalism, in analogy to recent work on self-diffusion near interfaces \cite{simonnin2017diffusion}, is another promising avenue. 
In this context, it also seems worthwhile to scrutinize a recent hypothesis according to which in concentrated electrolytes confined between two electrodes, the transport parameters are governed by volume rather than momentum conservation \cite{lorenz2022local}. 
Finally, our theoretical formalism likely also provides insights into the finite-size effects of ionic correlations. 
Recently, Shao et al. \cite{shao2020role} have shown numerically that while the diffusion coefficients show their well-known finite-size effects \cite{dunweg1993molecular2,yeh2004system,gabl2012computational} proportional to $L^{-1}$, the overall conductivity is independent of the system size, implying that the cross correlations must exhibit finite-size effects that compensate the finite-size effect of the diffusivity. 
Indeed, a finite-size effect proportional to $L^{-1}$ was found from their MD simulations for the cross correlations \cite{shao2020role}. 
Similar empirical observations have be made for mutual diffusivities in multicomponent systems \cite{jamali2018finite,celebi2021finite}. 
Jamali et al. \cite{jamali2018finite} found that the finite-size correction for Maxwell-Stefan diffusivities differs from that of the self-diffusion coefficients by a factor equal to the inverse thermodynamic factor. 
From Eq. \ref{eq:fsesh}, we recognize that the correction to the pair diffusion scales inversely with both the viscosity as well as the box length, similar to that of self-diffusion coefficients \cite{dunweg1993molecular2,yeh2004system,gabl2012computational}. 
As demonstrated by Jamali et al. \cite{jamali2018finite}, the impact of the structure of the liquid on the pair (or mutual) diffusion could be captured by the thermodynamic factors, which we leave for future analyses.

\begin{acknowledgments}
The authors thank Gerhard Hummer, Jens Smiatek and Volker Lesch for helpful discussions. 
\end{acknowledgments}

\section*{References}

\bibliography{refs}

\clearpage

\appendix

\renewcommand\thefigure{\thesection\arabic{figure}}
\setcounter{figure}{0}

\section{Numerical Evaluation of the Ewald Sum}
\label{sec:ewald}

\begin{figure}
 \centering
 \includegraphics[width=0.45\textwidth]{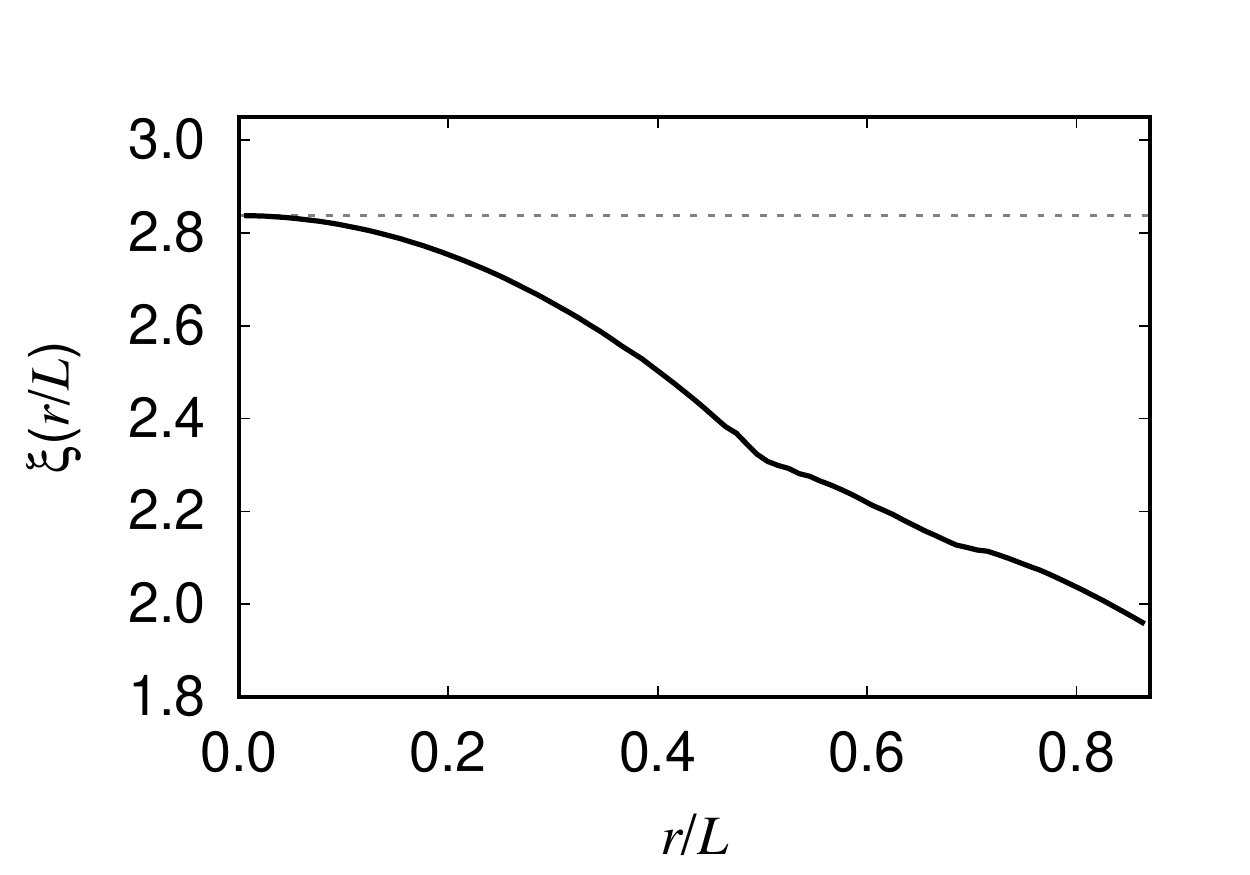}
\caption{\label{fig:xir} Distance dependence of the finite-size correction $\xi(r/L)$ in Eq. \ref{eq:fsesh}.}
\end{figure}

The distance dependence of $\xi(r/L)$ in Eq. \ref{eq:fsesh} was evaluated numerically (Figure \ref{fig:xir}). 
To this end, the convergence parameter $\alpha$ has been chosen such that both the summation in real space and in reciprocal space in Eq. \ref{eq:rfscnondim} converged with a reasonable number of lattice vectors ($30$ vectors in each spatial direction in our case), which in practice corresponds to a value of $\alpha L$ on the order of one. 
The orientation of the distance vector ${\bf r}$ has been sampled randomly and the lattice sums in Eq. \ref{eq:rfscnondim} have been carried out for all three dimensions. 
We note that for $r\rightarrow 0$, we recover the numerical value of $\xi \approx 2.837297$ reported previously \cite{dunweg1993molecular2,yeh2004system}. 
For larger distances, $\xi$ is a slowly varying function of $r$, and decays to about $70$ \% of its original value for $r/L=\sqrt{3}/2$. 
Slight kinks can be observed at $r/L = 1/2$ (maximum distance in one spatial direction) and $r/L = \sqrt{2}/2$ (maximum distance within a plane defined by any two spatial directions). 
Due to the fact that the above derivation approximates the particles as point-like, no further length scale like the particle radius enters the distance dependence shown in Figure \ref{fig:xir}. 

For the orientation-dependent flow fields $\Delta\mathbf{r}_\mathrm{PBC}(\hat{\mathbf r},r/L)$ in Figure \ref{fig:dtens}, we carried out analogous lattice sums, but retained the orientational dependence expressed by the tensor products. 
In particular, all real-space terms have been weighted by the tensor product $(\Delta \hat{\mathbf r}_0+(\hat{\mathbf r}\cdot\Delta\hat{\mathbf r}_0)\hat{\mathbf r})$, whereas the term evaluated in reciprocal space was weighted by $(\Delta\hat{\mathbf r}_0-(\hat{\mathbf k}\cdot\Delta\hat{\mathbf r})\hat{\mathbf k})$ (with $\Delta{\mathbf r}_0$ being the displacement vector of a particle in the center of the box: 
\begin{widetext}
 \begin{align}
 \begin{split}
  \label{eq:rfscnondimfu}
  \Delta{\mathbf r}_\mathrm{PBC}(\hat{\mathbf r},r/L) &= \frac{3}{4}\frac{k_\mathrm{B}T\Delta t}{\pi\eta L}\left[\left(\Delta\hat{\mathbf r}_0+(\hat{\mathbf r}\cdot\Delta\hat{\mathbf r}_0)\hat{\mathbf r}\right)\left(\frac{L}{r}\right) + \sum_{{\bf n} \neq {\bf 0}}\frac{2\left(\Delta\hat{\mathbf r}_0-(\hat{\mathbf k}\cdot\Delta\hat{\mathbf r}_0)\hat{\mathbf k}\right)}{\pi n^2}\exp{\left(2 \pi i\,{\bf n}\cdot({\bf r}/L)\right)}\exp{\left(-\frac{\pi^2 n^2}{(\alpha L)^2}\right)}\right. \\
   &+ \left.\left(\Delta\hat{\mathbf r}_0+(\hat{\mathbf r}\cdot\Delta\hat{\mathbf r}_0)\hat{\mathbf r}\right)\left(\sum_{{\bf n}\neq {\bf 0}}\frac{\erfc{\left((\alpha L)|({\bf r}/L) + {\bf n}|\right)}}{|({\bf r}/L) + {\bf n}|} - \frac{\erf{\left((\alpha L)(r/L)\right)}}{r/L} - \frac{\pi^2}{(\alpha L)^2}\right)\right]\text{.}
 \end{split}
 \end{align}
\end{widetext}
Unit values have been chosen for $k_\mathrm{B}T$, $\Delta t$, $\eta$ and $L$ in Figure \ref{fig:dtens}, and all vectors have been normalized to unit vectors.

\section{Comparison with [EMIm][BF$_4$]}
\label{sec:emimbf4}

\begin{figure*}
 \subfloat[\label{fig:rhor_eb_100ps}]{
 \includegraphics[width=0.5\textwidth]{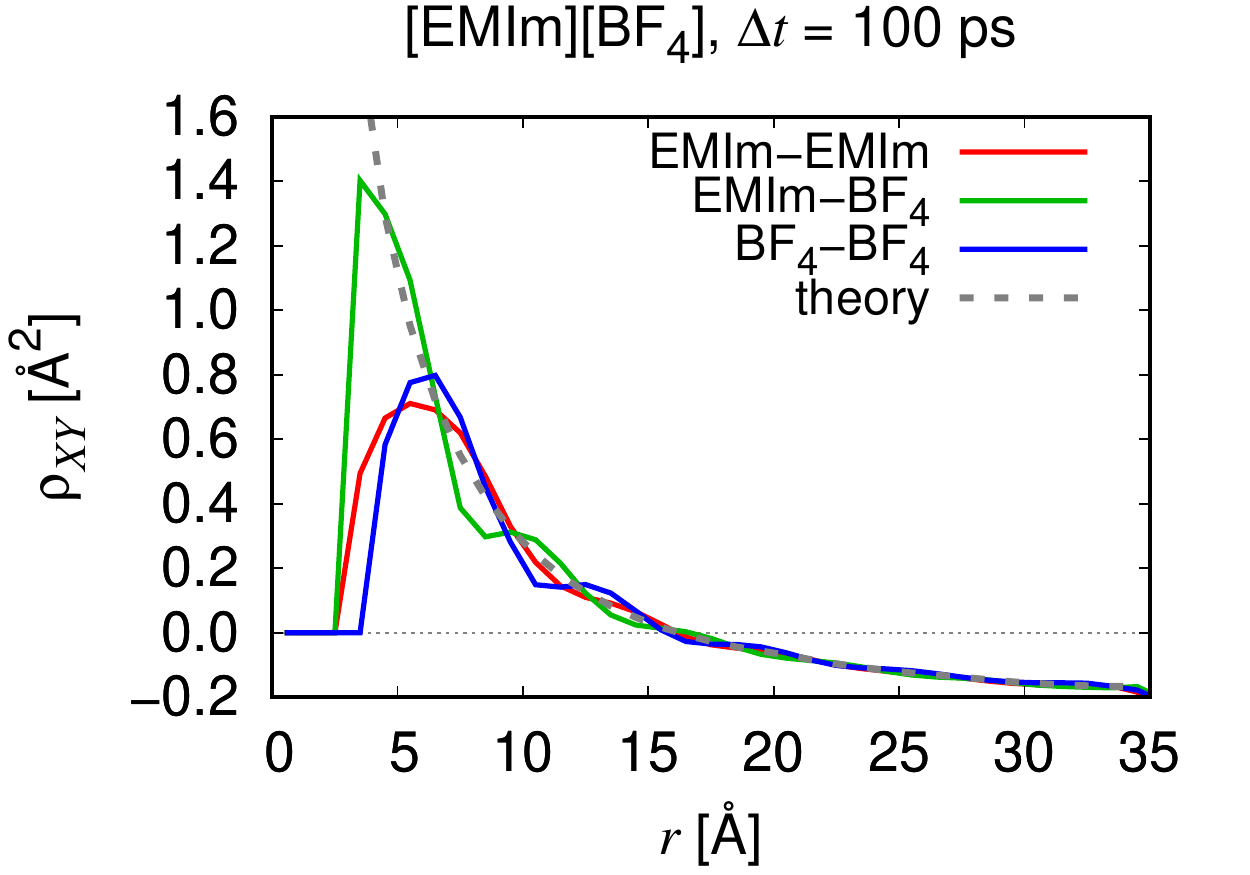}}
 \subfloat[\label{fig:drhor_eb_100ps}]{
 \includegraphics[width=0.5\textwidth]{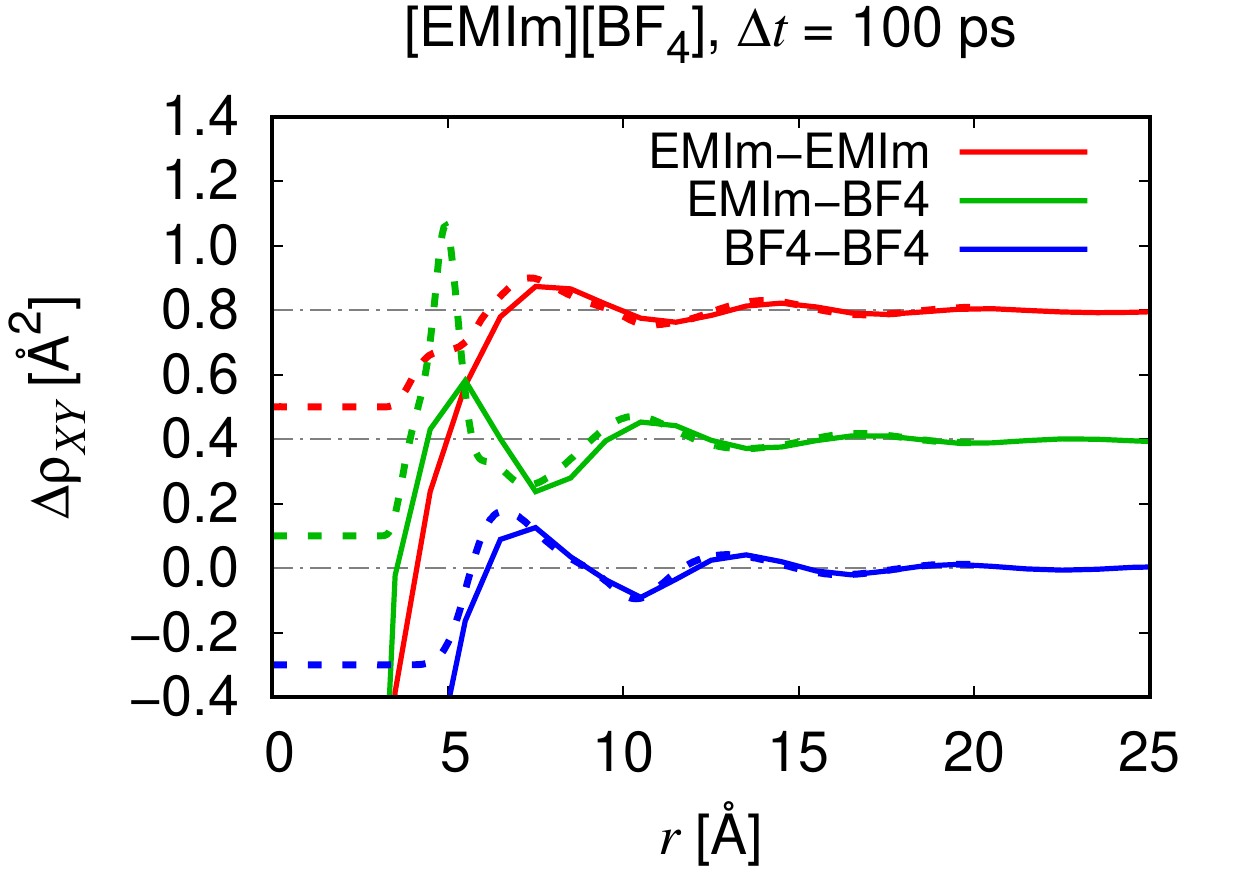}}
\\
 \subfloat[\label{fig:rhor-vs-t_eb}]{
 \includegraphics[width=0.5\textwidth]{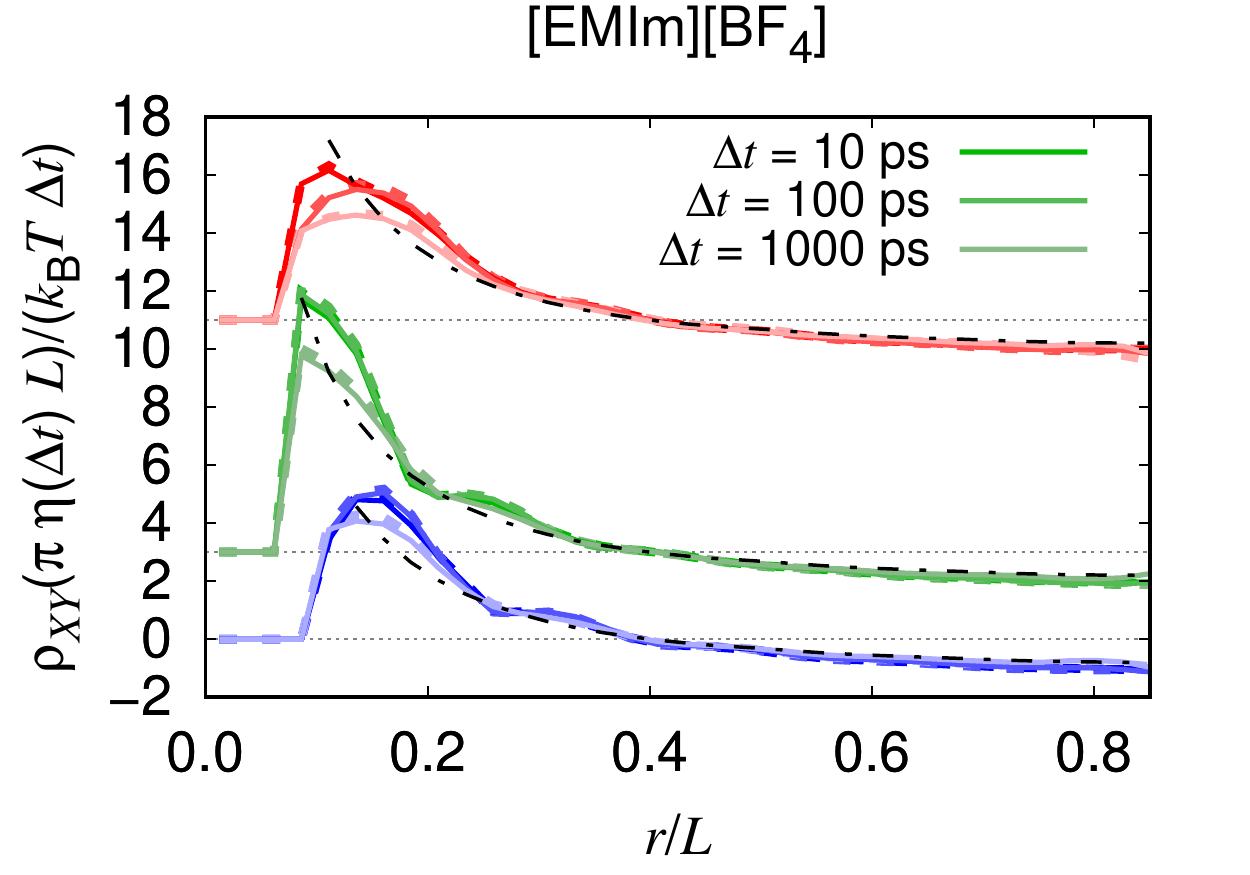}}
 \subfloat[\label{fig:4pir2rhor-vs-t_eb}]{
 \includegraphics[width=0.5\textwidth]{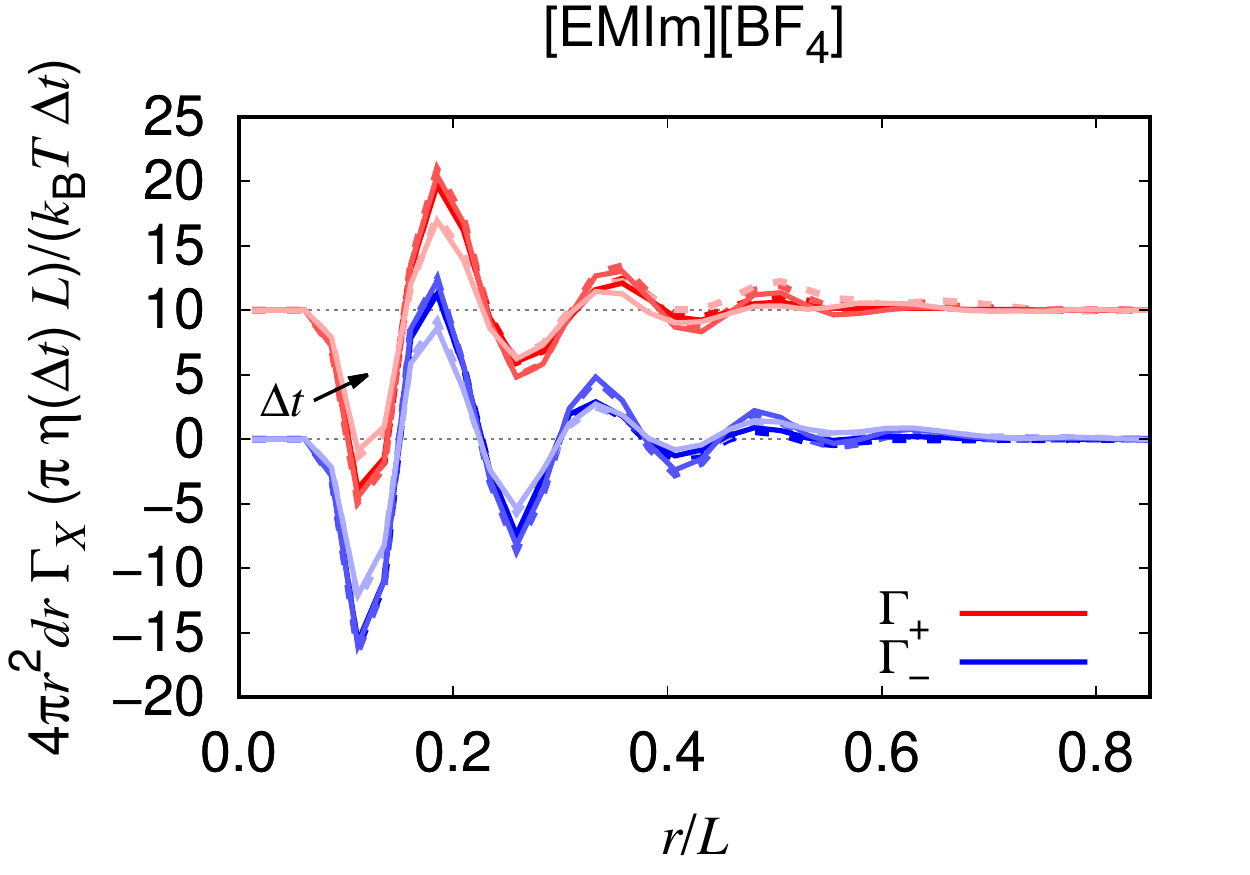}}
\\
 \subfloat[\label{fig:chir-vs-t_eb}]{
 \includegraphics[width=0.5\textwidth]{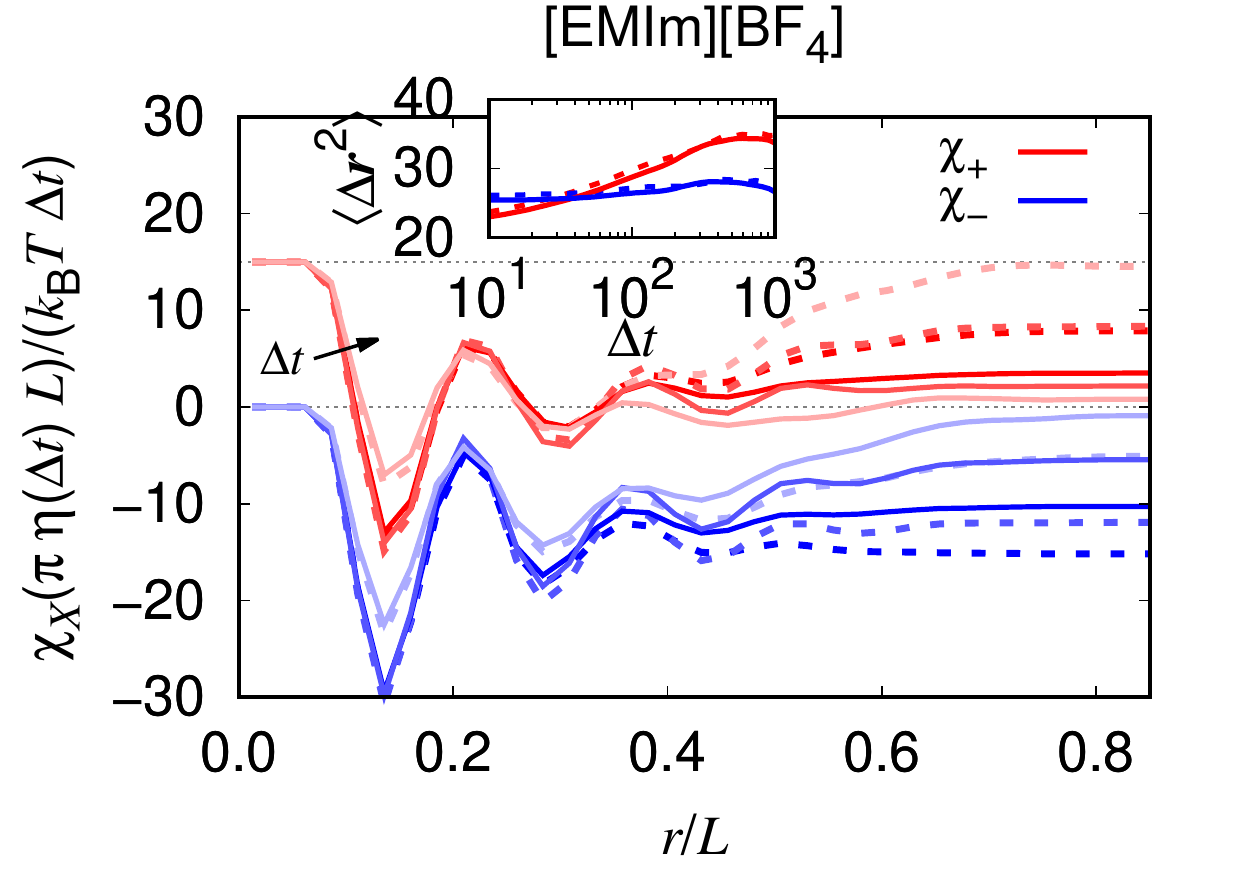}}
 \subfloat[\label{fig:sig-vs-t_eb}]{
 \includegraphics[width=0.5\textwidth]{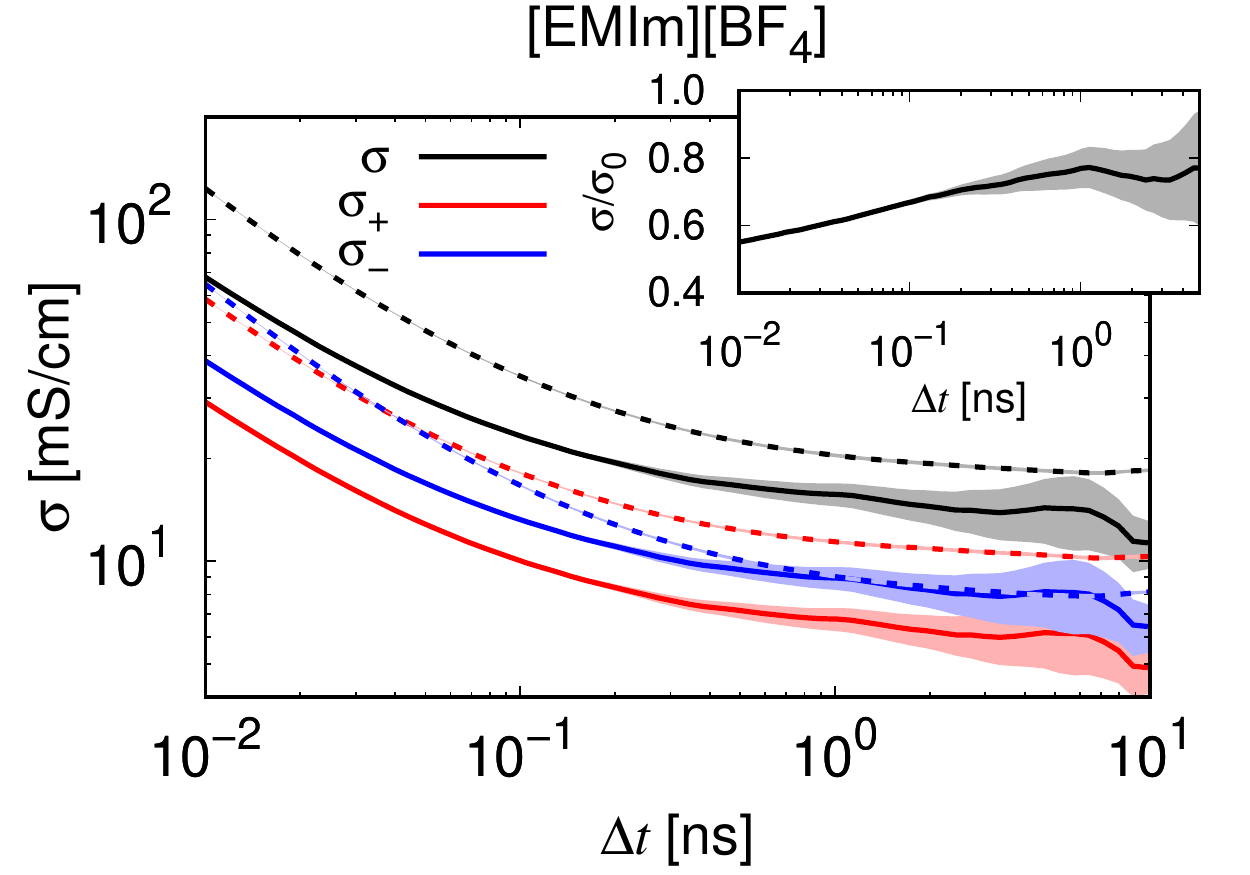}}
\caption{\label{fig:rhoreb} Results for [EMIm][BF$_4$]. (a): Distance-dependent correlation $\rho_{XY}(r)$ together with hydrodynamic fit (dashed), (b): Deviations of the simulation data from the hydrodynamic fit (solid) with the scaled and shifted RDFs (dashed), (c): Normalized $\rho_{XY}$ at different $\Delta t$ as a function of normalized distance $r/L$ together with the theory curve (dash-dotted), (d): Weighted and normalized cross-correlations $4\pi r^2 \Gamma_\pm$, (e): Integrated and normalized cross-correlations $\chi_\pm$, (f): Time dependent conductivities (solid) and ideal Nernst-Einstein conductivities (dashed). Several curves have been shifted for better visibility.}
\end{figure*}

In addition to [EMIm][TFSI], the IL [EMIm][BF$_4$] has been simulated as well. 
The analogous results are summarized in Figure \ref{fig:rhoreb}. 
We observe that the results are qualitatively the same as for [EMIm][TFSI].

\section{Viscosity Calculation}
\label{sec:visc}

\begin{figure*}
 \subfloat[]{
 \includegraphics[width=0.5\textwidth]{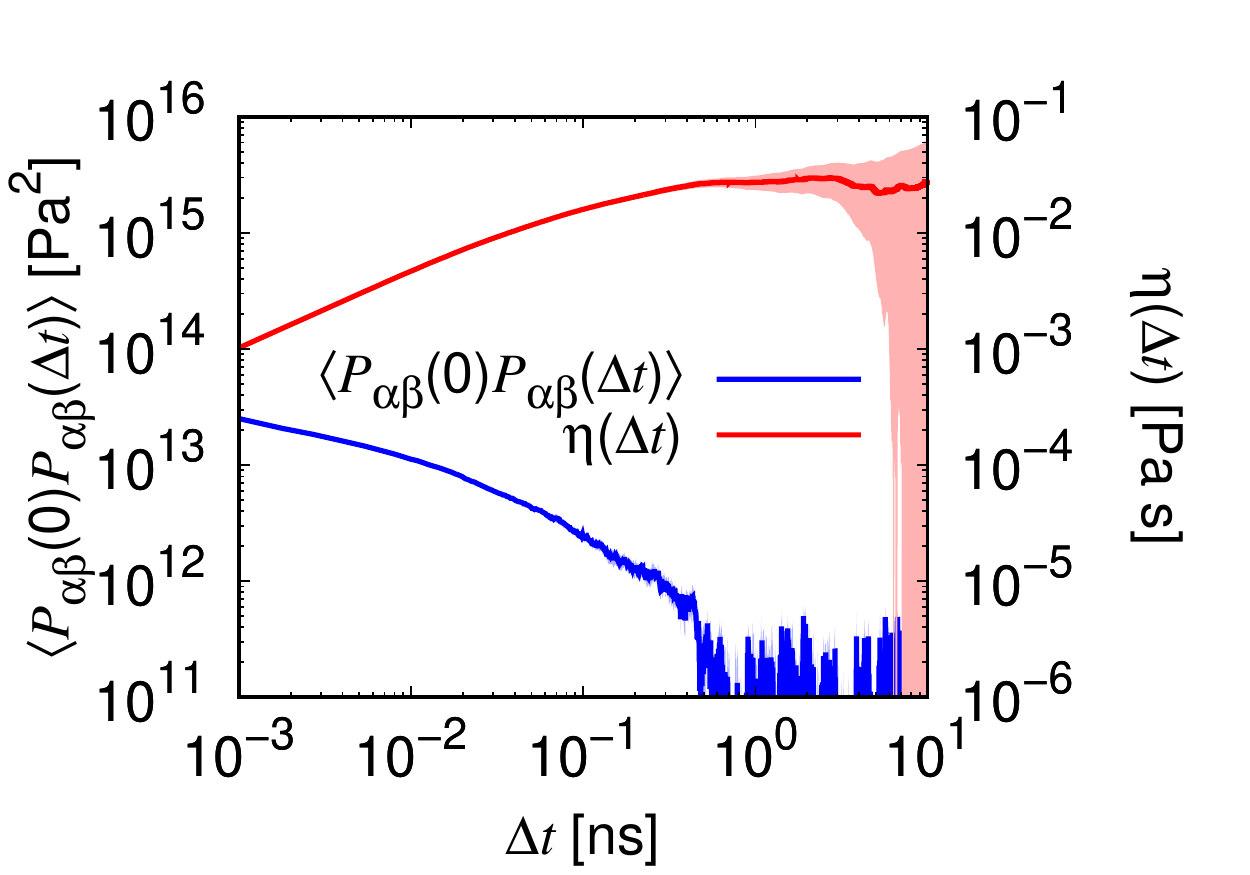}}
 \subfloat[]{
 \includegraphics[width=0.5\textwidth]{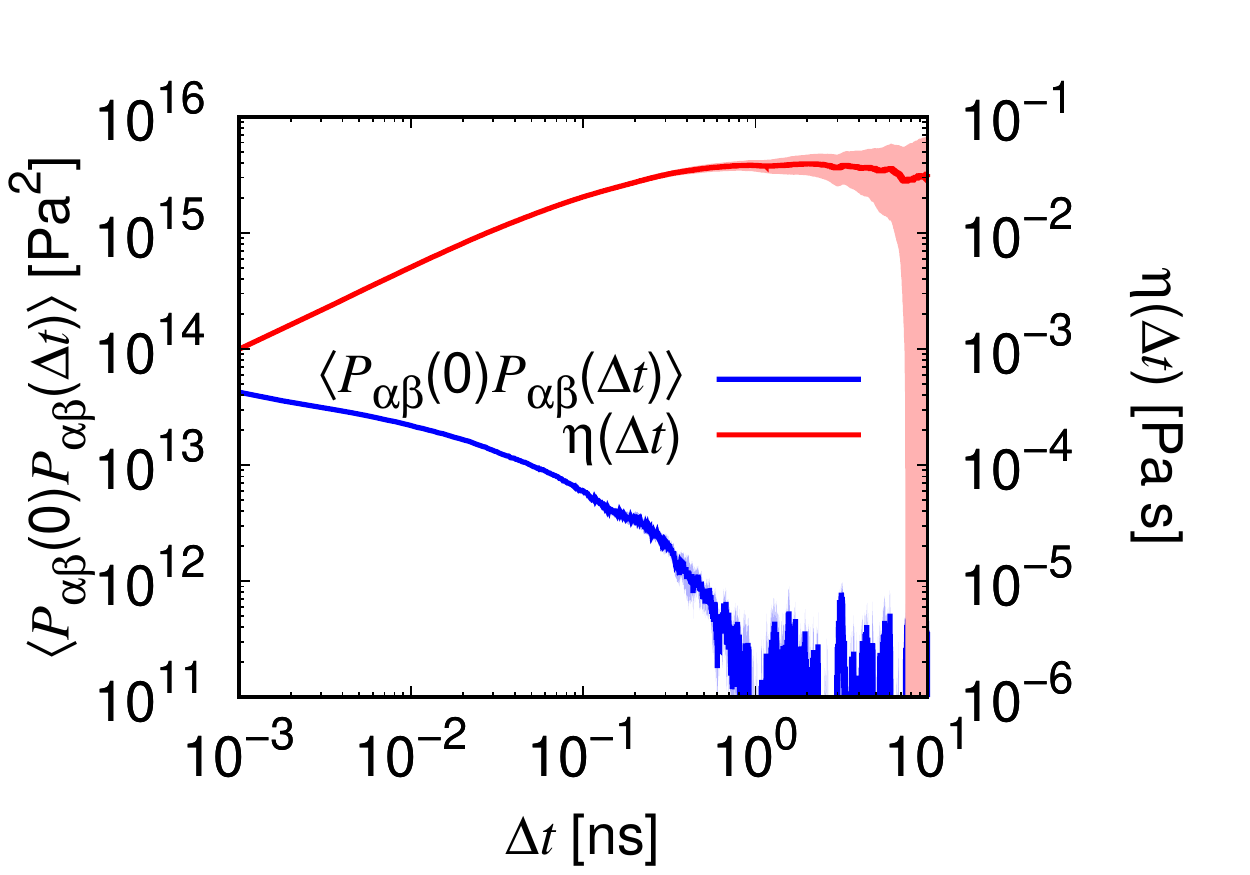}}
\\
 \subfloat[]{
 \includegraphics[width=0.5\textwidth]{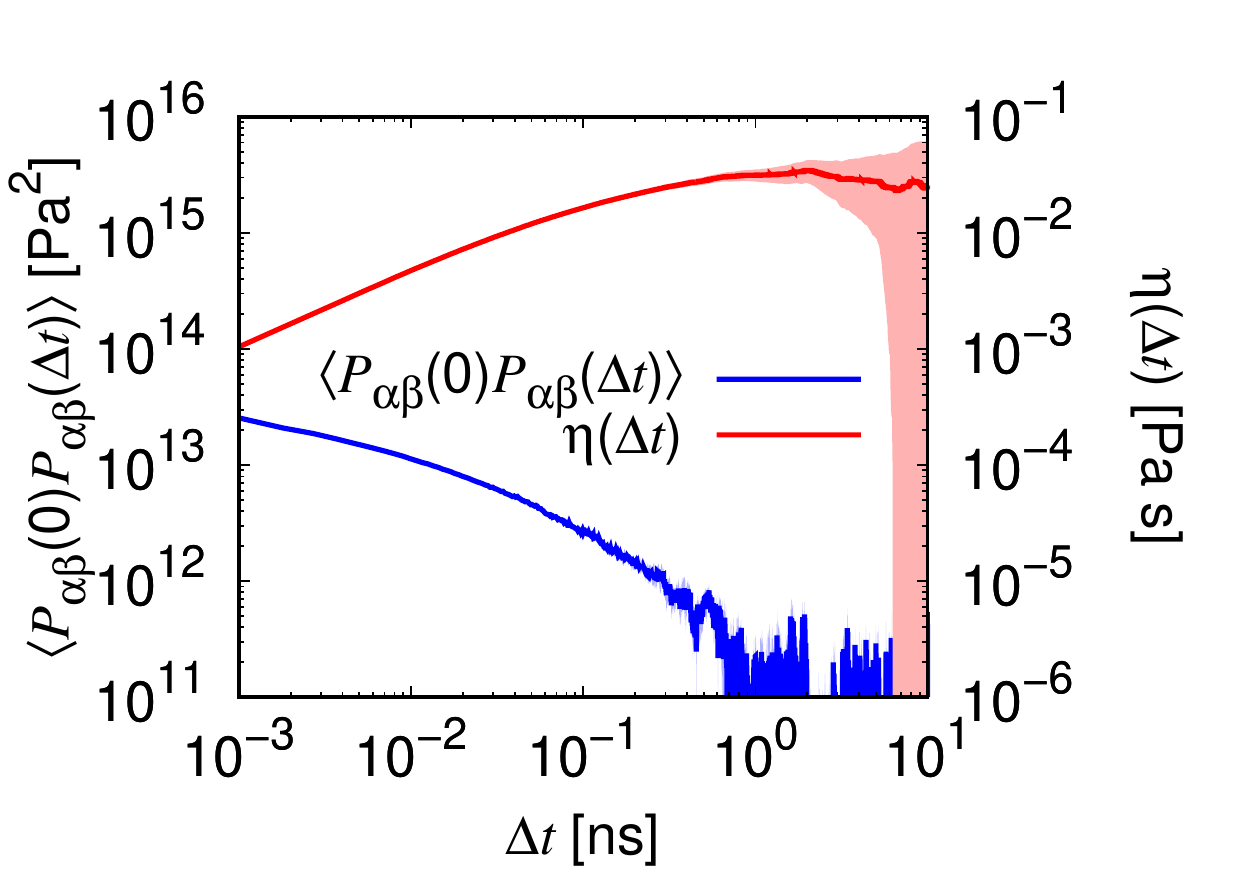}}
 \subfloat[]{
 \includegraphics[width=0.5\textwidth]{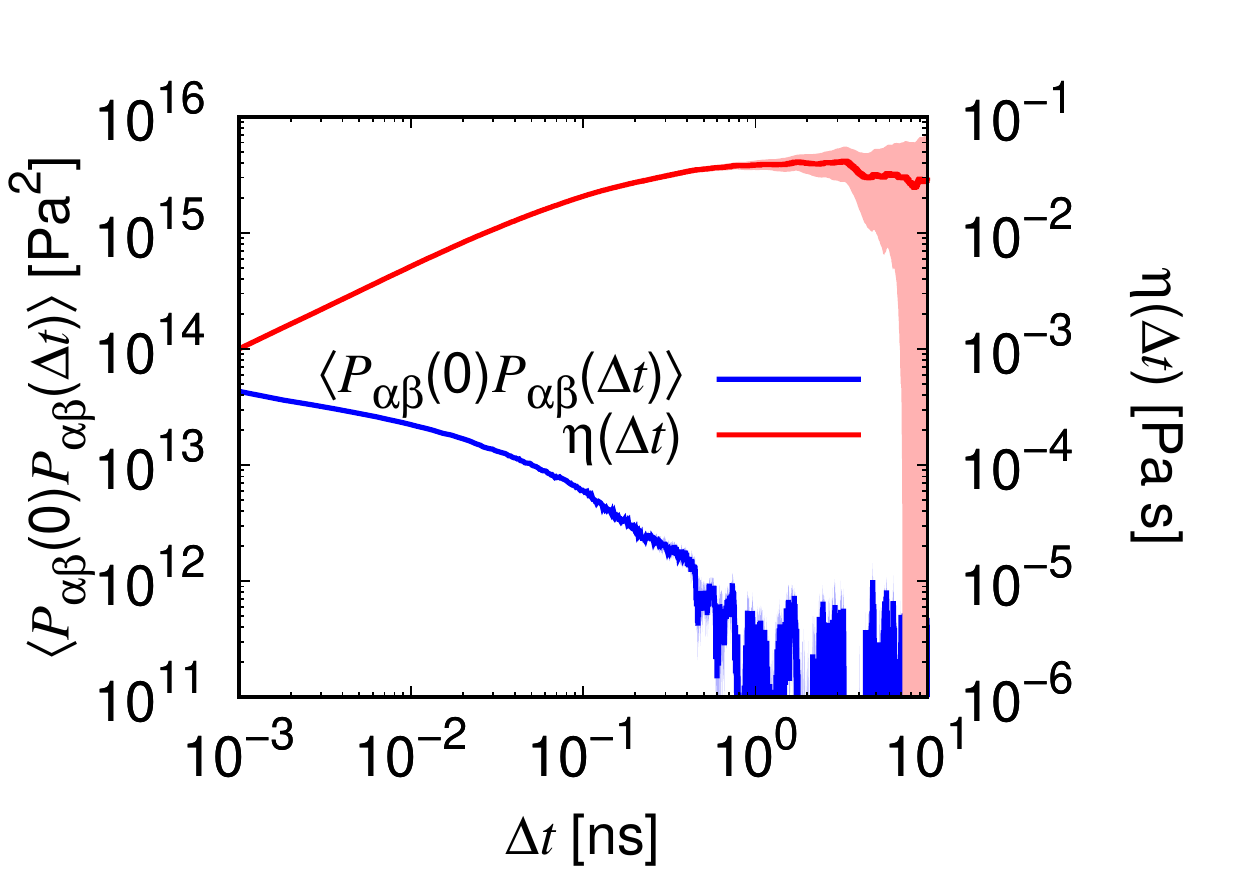}}
\\
 \subfloat[]{
 \includegraphics[width=0.5\textwidth]{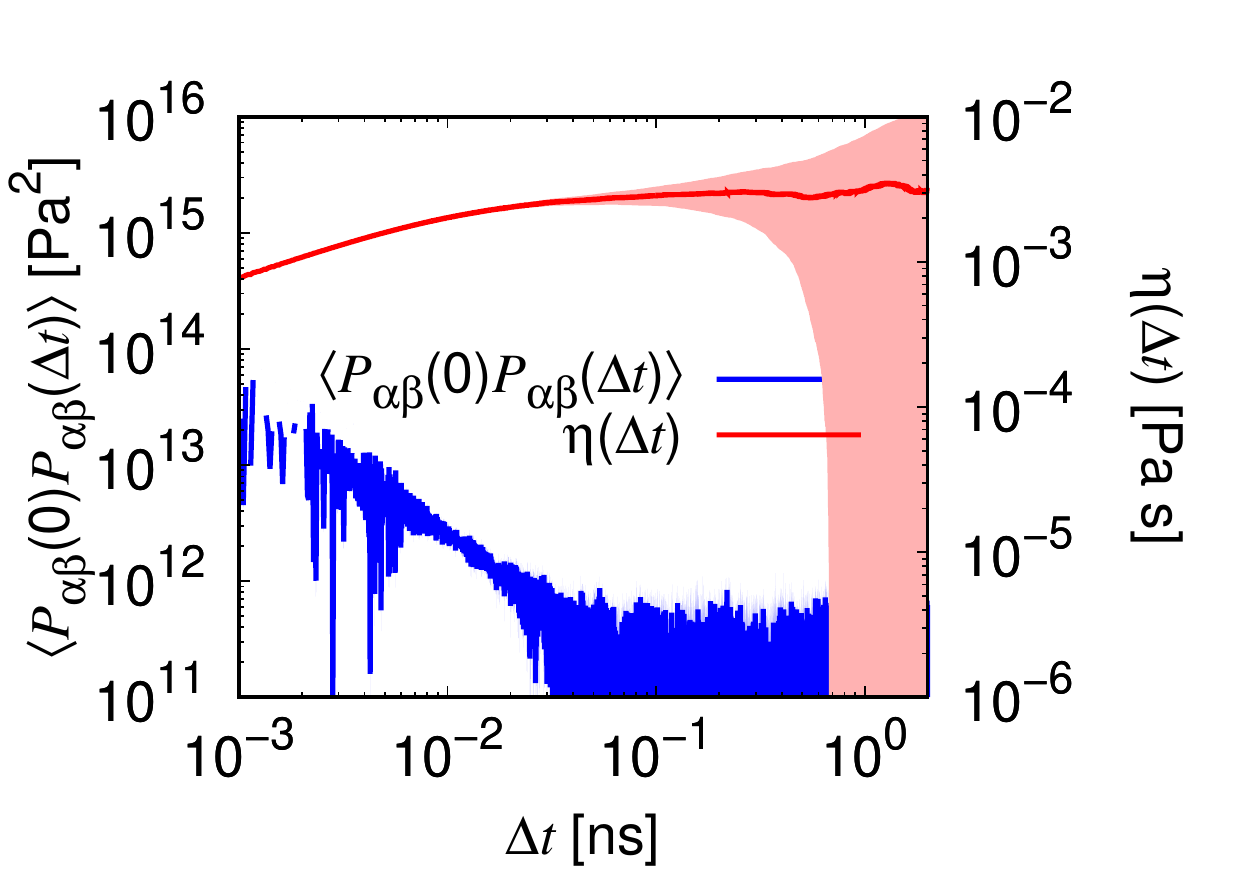}}
\caption{\label{fig:eta} Pressure tensor autocorrelation function $\langle P_{\beta\gamma}(0)P_{\beta\gamma}(t)\rangle$ and time-dependent viscosity $\eta(t)$ determined from the cumulative integral in Eq. \ref{eq:gketat} for (a) [EMIm][TFSI] and (b) [EMIm][BF$_4$] with ordinary masses, (c) [EMIm][TFSI] and (d) [EMIm][BF$_4$] with modified masses (section \ref{sec:sim}) as well as (e) EC/DMC/LiTFSI.}
\end{figure*}

To compute the prefactor in Eq. \ref{eq:fsesh}, the viscosity was extracted from the MD data via the autocorrelation function of the pressure tensor \cite{holian1983shear,yeh2004system} (Eqs. \ref{eq:gketa} and \ref{eq:gketat}). 
Figure \ref{fig:eta} shows $\langle P_{\beta\gamma}(0)P_{\beta\gamma}(\Delta t)\rangle$ and the corresponding integral according to the Green-Kubo relation in Eq. \ref{eq:gketat}. 
We note that for the ILs starting from around $500$ ps, the statistics deteriorates. 
Nonetheless, since the absolute value of $\langle P_{\beta\gamma}(0)P_{\beta\gamma}(\Delta t)\rangle$ is small for $\Delta t>500$ ps, the uncertainties of the corresponding cumulative integrals (shaded areas in Figure \ref{fig:eta}) are acceptable until a few nanoseconds, at which $\eta(\Delta t)$ converges to its long-time value. 
The long-time viscosity was estimated at $\Delta t = 1$ ns for both ILs, where the integral of Eq. \ref{eq:gketat} saturates. 
We obtain $\eta = 27.3 \pm 3.9$ mPa s for [EMIm][TFSI] and $\eta = 38.1 \pm 4.4$ mPa s for [EMIm][BF$_4$] at $\Delta t = 1$ ns. 
For the ILs with modified masses, we find corresponding values of $31.5 \pm 3.8$ and $38.4 \pm 3.6$ mPa s for [EMIm][TFSI] and [EMIm][BF$_4$], respectively. 
For larger $\Delta t$-values no further trend for $\eta(\Delta t)$ can be identified within the uncertainties. 
Remarkably, the viscosities for the ILs with modified masses are identical to those of the standard ILs, showing that the potential (mainly electrostatic) terms in Eq. \ref{eq:ptens} outweight the kinetic terms. 
For the CE, the statistics deteriorates from about $\Delta t>100$ ps. 
The estimated long-time viscosity at $\Delta t=300$ ps is $3.0 \pm 1.3$ mPa s. 

\end{document}